\documentclass[12pt]{article}
\usepackage{amsmath,amssymb,bbm,amsthm}
\usepackage{graphicx}
\usepackage{subcaption}
\usepackage{natbib}
\usepackage{url} 
\usepackage{xcolor}

\usepackage{algorithm}

\usepackage{multicol,multirow}

\newcommand{\blind}{0}

\newcommand{\EE}{\mathbb{E}}
\newcommand{\PP}{\mathbb{P}}
\newcommand{\Normal}{\text{N}}

\newcommand{\R}{\mathbb{R}}
\newcommand{\N}{\mathbb{N}}
\newcommand{\J}{\mathcal{J}}
\newcommand{\M}{\mathcal{M}}
\newcommand{\Q}{\mathcal{Q}}
\newcommand{\GP}{\mathcal{GP}}
\newcommand{\norm}[1]{\left\lVert#1\right\rVert}
\newcommand{\abs}[1]{\left|#1\right|}
\newcommand{\bracevec}[1]{\left\lbrace #1 \right\rbrace}
\newcommand{\iid}{\overset{iid}{\sim}}
\newcommand{\ind}{\overset{ind}{\sim}}

\newcommand{\Zmat}{\mb{Z}}
\newcommand{\parambigi}{w}
\newcommand{\parambig}{\mb{\parambigi}}
\newcommand{\parambigS}{\parambig^{*}}
\newcommand{\Wdim}{m}
\newcommand{\Udim}{d}
\newcommand{\addpredi}{\eta}
\newcommand{\addpred}{\mb{\addpredi}}
\newcommand{\addpredS}{\addpred^{*}}
\newcommand{\addpredSmode}{\widehat{\addpred}^{*}}

\newcommand{\llhessW}{\mb{C}_{\parambig}}
\newcommand{\llhessWS}{\mb{C}_{\parambigS}}
\newcommand{\llhesseta}{\mb{C}_{\addpred}}
\newcommand{\lphessWraw}{\mb{H}_{\parambig}}
\newcommand{\lphessW}{\mb{H}_{\Wmode_{\paramsmall}}}
\newcommand{\lphessWS}{\mb{H}^{*}_{\WmodeS_{\paramsmall}}}

\newcommand{\paramsmall}{\mb{\theta}}
\newcommand{\paramsmalldim}{s}
\newcommand{\paramsmalllik}{\paramsmall_{1}}
\newcommand{\paramsmallprior}{\paramsmall_{2}}
\newcommand{\paramsmalllikdim}{\paramsmalldim_{1}}
\newcommand{\paramsmallpriordim}{\paramsmalldim_{2}}
\newcommand{\covxi}{x}
\newcommand{\covx}{\mb{\covxi}}
\newcommand{\covxdim}{p}
\newcommand{\covzi}{z}
\newcommand{\covz}{\mb{\covzi}}
\newcommand{\covzdim}{r}
\newcommand{\etadim}{N}
\newcommand{\basisfunction}{\phi}
\newcommand{\uui}{u}
\newcommand{\uu}{\mb{\uui}}
\newcommand{\uprec}{\mb{Q}}
\newcommand{\uprecS}{\uprec^{*}}
\newcommand{\bb}{\mb{\beta}}
\newcommand{\betacov}{\mb{\Sigma}_{\bb}}
\newcommand{\tpose}{^{\texttt{T}}}

\newcommand{\zero}{\mb{0}}
\newcommand{\loc}{\mb{s}}
\newcommand{\Matern}{\text{M}}

\newcommand{\Wmode}{\widehat{\parambig}}
\newcommand{\WmodeS}{\widehat{\parambig}^{*}}

\newcommand{\quadpoint}{\mb{z}}

\newcommand{\quadpointset}{\mathcal{Q}}

\newcommand{\weight}{\omega}
\newcommand{\quadnum}{k}

\newcommand{\I}{\mb{I}}
\newcommand{\datai}{y}
\newcommand{\data}{\mb{\datai}}
\newcommand{\approxpi}{\widetilde{\pi}}

\newcommand{\G}{\text{\tiny{\tt G}}}
\newcommand{\LA}{\text{\tiny{\tt LA}}}

\newcommand{\gaussapprox}{\widetilde{\pi}_{\G}}
\newcommand{\laplaceapprox}{\widetilde{\pi}_{\LA}}
\newcommand{\laplacemode}{\widehat{\paramsmall}_{\LA}}
\newcommand{\laplacehess}{\widehat{\mb{H}}_{\LA}}
\newcommand{\laplacechol}{\widehat{\mb{L}}_{\LA}}

\newcommand{\paramsmalltrue}{\paramsmall_{0}}
\newcommand{\parambigtrue}{\parambig_{0}}

\newcommand{\mb}[1]{\boldsymbol{#1}}

\newcommand{\ELGM}{\texttt{ELGM}}

\newcommand{\RINLA}{\texttt{R-INLA}}
\newcommand{\MCMC}{\texttt{MCMC}}
\newcommand{\rstan}{\texttt{rstan}}
\newcommand{\TV}{\text{\texttt{TV}}}

\newcommand{\tvnorm}[1]{\norm{#1}_{\TV}}

\newcommand{\borelset}{\mathcal{K}}
\newcommand{\borelsets}[1]{\mathcal{B}(\R^{#1})}

\def\[#1\]{\begin{equation}\begin{aligned}#1\end{aligned}\end{equation}}
\def\*[#1\]{\begin{align*}#1\end{align*}}

\newtheorem{theorem}{Theorem}

\begin{document}

\def\spacingset#1{\renewcommand{\baselinestretch}%
{#1}\small\normalsize} \spacingset{1}


\if0\blind
{
  \title{\bf Fast, Scalable Approximations to Posterior Distributions in Extended Latent Gaussian Models}
  \author{Alex Stringer\thanks{
    The authors gratefully acknowledge funding from the Natural Sciences and Engineering Research Council of Canada.}\hspace{.2cm}\\
    Department of Statistics and Actuarial Science, University of Waterloo\\
    and \\
    Patrick Brown \\
    Department of Statistical Sciences, University of Toronto \\
    Centre for Global Health Research \\
    and \\
    Jamie Stafford \\
    Department of Statistical Sciences, University of Toronto}
  \date{}
  \maketitle
} \fi

\if1\blind
{
  \bigskip
  \bigskip
  \bigskip
  \begin{center}
    {\LARGE\bf Title}
\end{center}
  \medskip
} \fi

\bigskip
\begin{abstract}
We define a novel class of additive models, called Extended Latent Gaussian Models, that allow for a wide range of response distributions and flexible relationships between the additive predictor and mean response. The new class covers a broad range of interesting models including multi-resolution spatial processes, partial likelihood-based survival models, and multivariate measurement error models. Because computation of the exact posterior distribution is infeasible, we develop a fast, scalable approximate Bayesian inference methodology for this class based on nested Gaussian, Laplace, and adaptive quadrature approximations. We prove that the error in these approximate posteriors is $o_{p}(1)$ under standard conditions, and provide numerical evidence suggesting that our method runs faster and scales to larger datasets than methods based on Integrated Nested Laplace Approximations and Markov Chain Monte Carlo, with comparable accuracy. We apply the new method to the mapping of malaria incidence rates in continuous space using aggregated data, mapping leukaemia survival hazards using a Cox Proportional-Hazards model with a continuously-varying spatial process, and estimating the mass of the Milky Way Galaxy using noisy multivariate measurements of the positions and velocities of star clusters in its orbit.
\end{abstract}

\noindent%
\vfill

\newpage
\spacingset{1.5} 


\section{Introduction}\label{sec:intro}

\subsection{Latent Gaussian and Extended Latent Gaussian Models}

Latent Gaussian Models (LGMs; \citealt{inla}) are a class of additive regression models that are of broad interest in modern practice. The class of LGMs includes generalized linear models \citep{bayesianglmm}, generalized additive models \citep{faststable}, semi-parametric \citep{inlasurvival} and joint \citep{competingrisksjointmodelsinla} models for survival data, and spatial models \citep{spde,inlasoftware}. However, a central assumption of the LGM (\ref{eqn:lgm}) is that the response variables $\datai_{i}$ are conditionally independent, given a single natural parameter $\addpredi_{i}$ each, or equivalently, that for each $i\in[n]$, $\mu_{i} = \EE(\datai_i|\addpredi_i)$ is a function of only $\eta_{i}$. This limits the breadth of models that belong to the class of LGMs.

The contributions of this paper are: (a) a class of models is introduced, called Extended Latent Gaussian Models (ELGMs), that permit much more general dependence between the $\mu_{i}$ and the natural parameters (\S\ref{subsec:elgms:definition}), (b) a fast and scalable algorithm is developed for making approximate Bayesian inferences in complex ELGMs and with large data sets (\S\ref{subsec:elgms:computational}), and (c) convergence of the approximate posteriors is established (Theorem \ref{thm:convergence} in \S\ref{sec:theory}). The new class of ELGMs includes multi-resolution spatial point processes \citep{disaggregation,aggspatial,geostataggdata}, partial likelihood-based survival \citep{partialspatial} and multinomial logistic \citep{stringer} models, zero-inflated models \citep{zeroinflated} including those for spatially-correlated data \citep{geostatlowresource}, and models incorporating multivariate measurement errors \citep{gwen3}. 

\subsection{Existing Methods for Latent Gaussian Models}

Bayesian inferences for LGMs are based upon a posterior distribution which is intractable, and  approximate inference algorithms mitigate this challenge. A number of Markov Chain Monte Carlo (MCMC) algorithms are available, including those based on block-updating \citep{blockupdate} and elliptical slice sampling \citep{ellipticalslicesampling}, as well as the general algorithms available in the STAN language \citep{stan}. Beyond MCMC, deterministic approximation methods exist that offer speed and scalability to large data. The Integrated Nested Laplace Approximation (INLA; \citet{inla,inlasoftware,bayesiancomputinginlanewfeatures,inlareview}) is one such method based on Gaussian, Laplace, and quadrature approximations. Variational inference \citep{variationalinference} is another such method, and has been applied in models for which INLA cannot be used \citep{ding,seth}.

The use of Laplace and quadrature approximations in fitting generalized linear mixed models is well precedented \citep{pql,approximationsnlme}. Recent approximation algorithms for generalized additive models make use of Gaussian approximations \citep{smoothestimation}. Taken separately, these posterior approximations each have established convergence theory \citep{walker,higherorderlaplace,adaptive}. However, as described in \S\ref{sec:prelims} of this paper, INLA is based on a nested application of these three types of approximations. Consequently, no formal convergence results have been derived for INLA. The method we propose in \S\ref{sec:elgms} is based on a nested approximation strategy similar to INLA, but applies to a richer class of models (\S\ref{sec:elgms}), scales well with increasing sample size (\S\ref{sec:comp}), and has approximation error shown to be $o_{p}(1)$ under standard conditions (Theorem \ref{thm:convergence}, \S\ref{sec:theory}; Online Supplement A).

\subsection{Computation For Extended Latent Gaussian Models}\label{subsec:intro:computational}

An LGM (Equation \ref{eqn:lgm} in \S\ref{sec:prelims}) depends on an additive predictor or natural parameter $\addpred = \Zmat\parambig\in\R^{n}$, where $\parambig\in\R^{\Wdim}$ are parameters, and $\Zmat\in\R^{n\times\Wdim}$ is a known design matrix. The likelihood may depend on additional parameters $\paramsmalllik$ (\S\ref{sec:prelims}). Inferences for $\parambig$ and $\paramsmalllik$ based on Gaussian and Laplace approximations require evaluation of the (negative) Hessian matrix, $\llhessW(\parambig,\paramsmalllik)\in\R^{\Wdim\times\Wdim}$ of the log-likelihood $\pi(\mb{Y}|\addpred,\paramsmalllik) = \prod_{i=1}^{n}\pi(Y_{i}|\addpredi_{i},\paramsmalllik)$. However, INLA employs a further modification of the LGM \citep{inla}, basing inference off of the noisy parameter $\parambigS = (\addpredS,\parambig)$, where $\addpredS\sim\Normal(\addpred,\tau^{-1}\I_{n})$ for some large, fixed $\tau>0$. Consequently, the Hessian matrix that INLA must evaluate to make approximate inferences is $\llhessWS(\parambigS,\paramsmalllik)\in\R^{(\Wdim+n)\times(\Wdim+n)}$, a larger matrix whose size increases linearly with the sample size, $n$, irrespective of the properties of the model being fit.

The conditional independence of $\datai_{i}$ given $\addpredi_{i}$ that is required in an LGM forces the upper left $n\times n$ block of $\llhessWS(\parambigS,\paramsmalllik)$ to be diagonal, and the use of algorithms for efficient decomposition of sparse matrices \citep{fastsamplinggmrf} make computations involving this matrix feasible for LGMs fit to moderately sized data. In contrast, in the ELGMs we introduce in \S\ref{sec:elgms}, this conditional independence is relaxed, and the upper left $n\times n$ block of $\llhessWS(\parambigS,\paramsmalllik)$ is no longer diagonal and may, in fact, be dense. In \S\ref{subsec:elgms:computational} we present a computational strategy for approximate inference in ELGMs that does not make use of INLA's noisy additive predictor. This leads to matrices which may be denser in the case of LGMs, but, cruicially, are always smaller by $n$ dimensions than the corresponding matrices that would be required if implementing INLA's approximations for ELGMs. The substantial performance gains of our strategy, especially at large sample sizes, are demonstrated in \S\ref{sec:comp}.

\subsection{Plan of Paper}

This paper is organized as follows. In \S\ref{sec:prelims} we describe existing approaches to approximate inference in LGMs. In \S\ref{sec:elgms} we introduce the class of ELGMs and an algorithm (Algorithm \ref{alg:implementation}) for making approximate Bayesian inferences based on them. In \S\ref{sec:theory} we present a formal convergence result for our posterior approximation in the form of Theorem \ref{thm:convergence}, and discuss model-specific theoretical considerations. In \S\ref{sec:comp} we provide numerical evidence suggesting that our approximations are faster and scale to larger datasets than other methods, while remaining accurate.

In \S\ref{sec:examples}, we apply our new methodology to fit three challenging models which are beyond the class of LGMs but belong to the new class of ELGMs. We fit a fully Bayesian analysis of aggregated spatial point process data, a challenge which \cite{lgcpagg} and others address using data augmentation techniques and MCMC; the existence of software for doing so in this example enables direct comparison of our method with MCMC techniques where we make similar inferences at substantial reduction in computational cost. We fit a Cox proportional hazards model with partial likelihood and spatially-varying hazard; previous analyses of these data employ parametric \citep{leukaemia,spde} and semi-parametric \citep{inlasurvival} models for the hazard due to the difficulty of working with both the partial likelihood and the latent spatial process. Finally, we fit the Galactic Mass Estimation model of \citet{gwen3} and \citet{gwen4} for estimating the mass of the Milky Way galaxy in the presence of multivariate measurement errors. Implementation makes use of the publicly available \texttt{aghq} package in the \texttt{R} language, and code for all examples is made available at \url{https://github.com/awstringer1/elgm-paper-code}.

\section{Preliminaries}\label{sec:prelims}

\subsection{Notation and Definitions}\label{subsec:prelim:notation}

In what follows, we use the notation $[n] = \left\{1,\ldots,n\right\}$, denote (column) vectors by lower-case $\mb{v} = (v_{i})_{i\in[n]}\in\R^{n}$, and sub-vectors by $\mb{v}_{S} = (v_{i})_{i\in S}$ for any $S\subseteq[n]$. We denote the concatenation of vectors $\mb{v}_{1},\ldots,\mb{v}_{p}$, having dimensions $d_{1},\ldots,d_{p}$, by $\mb{v} = (\mb{v}_{j})_{j\in[p]} = (v_{11},\ldots,v_{pd_{p}})\in\R^{d_{1}+\cdots+d_{p}}$. We denote matrices by upper-case $\mb{Z} = (Z_{ij})_{i\in[n],j\in[d]}\in\R^{n\times d}$. An LGM takes the following form:
\begin{equation}\begin{aligned}\label{eqn:lgm}
\datai_{i}|\addpredi_{i},\paramsmalllik&\ind \pi(\datai_{i}|\addpredi_{i},\paramsmalllik), \ \mu_{i} = \EE(Y_{i} | \addpredi_{i}) = g(\addpredi_{i}), \ \addpredi_{i} = \mb{x}_{i}^{T}\mb{\beta} + \sum_{q=1}^{r}u_{q}(z_{iq}),
\end{aligned}\end{equation}
The response variable is $\data = (\datai_{i})_{i\in[n]}$ having likelihood $\pi(\data|\addpred,\paramsmalllik) = \prod_{i=1}^{n}\pi(\datai_{i}|\addpredi_{i},\paramsmalllik)$. Denote the conditional mean of each response by $\mu_{i} = \EE(\datai_{i}|\addpredi_{i}),i\in[n]$, and define an inverse link function $g:\R\to\R$ such that $\mu_{i} = g(\addpredi_{i})$. The parameters $\paramsmalllik\in\R^{\paramsmalllikdim}$, where $\paramsmalllikdim$ is assumed small, represent additional parameters of the likelihood, such as dispersion parameters. We denote the observed covariates, assumed fixed and known, by $\covx_{i}\in\R^{\covxdim}$ and $\covz_{i}\in\R^{\covzdim}$. 

Our treatment of the unknown functions $u_{q}(\cdot),q\in[r]$ follows that of \citet{smoothestimation}. We define a basis expansion $u_{q}(z) = \sum_{j=1}^{\Udim_{q}}\basisfunction_{qj}(z)\uui_{qj}$, where the $\basisfunction_{qj}(\cdot)$ are known basis functions (subject to identifiability constraints) and the $\uu_{q}=(\uui_{qj})_{j\in[\Udim_{q}]}$ are unknown weights to be inferred. Different choices of $\basisfunction_{qj}(\cdot)$ lead to different models, including: random effects models (\S\ref{sec:comp}), semi-parametric regression/generalized additive models, spatial models (\S\ref{subsec:aggspatial},\S\ref{subsec:coxph}), and measurement error models (\S\ref{subsec:astro}). The full vector of parameters (basis function weights) relating to $u_{q}(\cdot),q\in[\covzdim]$ is $\uu = (\uu_{q})_{q\in[\covzdim]}\in\R^{\Udim}$, with $\Udim = \sum_{q=1}^{\covzdim}\Udim_{q}$.

Define $\parambig = (\uu,\bb)\in\R^{\Wdim}, \Wdim = \Udim+\covxdim$, and write $\addpred = \Zmat\parambig$ as in \S\ref{sec:intro}, for design matrix $\Zmat\in\R^{n\times\Wdim}$ depending on $\basisfunction_{qj},\covz_{i},\covx_{i},i\in[n],j\in[\Udim],q\in[\covzdim]$. For the remainder, we will use both $\pi(\data|\addpred,\paramsmalllik)$ and $\pi(\data|\parambig,\paramsmalllik)$ to denote the likelihood, as context dictates. A Gaussian prior is placed on $\uu|\paramsmallprior\sim\Normal[\zero,\uprec_{\uu}^{-1}(\paramsmallprior)]$, where the known precision (inverse covariance) matrix $\uprec_{\uu}(\paramsmallprior)$ depends upon a further parameter $\paramsmallprior\in\R^{\paramsmallpriordim}$, where $\paramsmallpriordim$ is assumed small. A Gaussian prior $\bb\sim\Normal\left(\zero,\betacov\right)$ with fixed covariance $\betacov$, usually taken to be diagonal, is placed on $\bb$. Let $\uprec(\paramsmallprior) = \text{diag}\left\{ \uprec_{\uu}(\paramsmallprior),\betacov^{-1}\right\}$, so $\parambig|\paramsmallprior\sim\Normal[\zero,\uprec^{-1}(\paramsmallprior)]$. Further define $\paramsmall = (\paramsmalllik,\paramsmallprior)\in\R^{\paramsmalldim},\paramsmalldim = \paramsmalllikdim+\paramsmallpriordim$, with prior $\pi(\paramsmall)$.

\subsection{Posteriors and Approximate Inference}\label{subsec:prelim:approximateinference}

Inferences for the parameters of interest, $\parambig$ and $\paramsmall$, are based, in principle, on the posterior distributions
\begin{equation}\begin{aligned}
\pi(\parambig|\data) &= \int\pi(\parambig|\data,\paramsmall)\pi(\paramsmall|\data)d\paramsmall,\\
\pi(\paramsmall|\data) &= \int\pi(\parambig,\paramsmall|\data)d\parambig,
\end{aligned}\end{equation}
where
\begin{equation}
\pi(\parambig|\data,\paramsmall)\propto\exp\bracevec{-\frac{1}{2}\parambig\tpose\uprec(\paramsmallprior)\parambig + \log\pi(\data|\parambig,\paramsmalllik)}.
\end{equation}
All of these posteriors are intractable, and in practice, inferences are based on approximations. For fixed $\paramsmall$, define $\Wmode_{\paramsmall} = \text{argmax}_{\parambig}\log\pi(\parambig,\data,\paramsmall)$ and $\lphessW(\paramsmall) = -\partial^{2}_{\parambig}\log\pi(\Wmode_{\paramsmall},\data,\paramsmall) = \uprec(\paramsmallprior) + \llhessW(\Wmode_{\paramsmall},\paramsmalllik)$. A Gaussian approximation to $\pi(\parambig|\data,\paramsmall)$ is given by 
\begin{equation}\label{eqn:gaussapprox}
\gaussapprox(\parambig|\data,\paramsmall) = (2\pi)^{-\Wdim/2}\abs{\lphessW(\paramsmall)}^{1/2}\exp\bracevec{-\frac{1}{2}(\parambig-\Wmode_{\paramsmall})\tpose\lphessW(\paramsmall)(\parambig-\Wmode_{\paramsmall})}.
\end{equation} 
Define $\quadpointset(\paramsmalldim,\quadnum)$ to be the set of points, and $\weight_{\quadnum}:\quadpointset(\paramsmalldim,\quadnum)\to\R$ to be the weights, from a quadrature rule (to be specified) in $\paramsmalldim$ dimensions with $\quadnum$ points per dimension. Let $\laplaceapprox(\paramsmall,\data) = \pi(\Wmode_{\paramsmall},\paramsmall,\data)/\gaussapprox(\Wmode_{\paramsmall}|\data,\paramsmall)$, $\laplacemode = \text{argmax}_{\paramsmall}\log\laplaceapprox(\paramsmall,\data)$, $\laplacehess(\paramsmall) = -\partial^{2}\log\laplaceapprox(\paramsmall,\data)$, and $\laplacehess^{-1}(\laplacemode) = \laplacechol\laplacechol\tpose$ where $\laplacechol$ is the lower Cholesky triangle. A Laplace approximation to $\pi(\paramsmall|\data)$ is given by:
\begin{equation}\label{eqn:laplace}
\laplaceapprox(\paramsmall|\data) = \frac{\laplaceapprox(\paramsmall,\data)}{\abs{\laplacechol}\sum_{\quadpoint\in\quadpointset(\paramsmalldim,\quadnum)}\laplaceapprox(\laplacechol\quadpoint+\laplacemode,\data)\weight_{\quadnum}(\quadpoint)}.
\end{equation}
Finally, define the nested approximation
\begin{equation}\label{eqn:nested}
\approxpi(\parambig|\data) = \abs{\laplacechol}\sum_{\quadpoint\in\quadpointset(\paramsmalldim,\quadnum)}\gaussapprox(\parambig|\data,\laplacechol\quadpoint+\laplacemode)\laplaceapprox(\laplacechol\quadpoint+\laplacemode|\data)\weight_{\quadnum}(\quadpoint),
\end{equation}
where the same adapted points and weights have been used to approximate this second, different integral.

\citet{inla} make approximate inferences starting from (\ref{eqn:nested}), with three important differences. First, their INLA method focusses on approximating marginal posteriors $\approxpi(\parambigi_{j}|\data),j\in[\Wdim]$, for which an additional Laplace approximation is used, using a computational strategy recently expanded upon by \citet{simplifiedinla}. Our focus in the present work is on the joint posterior $\approxpi(\parambig|\data)$, as evidenced in the examples of \S\ref{sec:comp} and \S\ref{sec:examples}, so we do not pursue this here for ELGMs. Second is the choice of quadrature rule, for which they consider the empirical performance of several different options. As we demonstrate in the proof of Theorem \ref{thm:convergence} in Online Supplement A, the specific properties of adaptive (Gauss-Hermite) quadrature are used when proving convergence of the overall approximation (\ref{eqn:nested}), and it is not immediate that similar properties hold for other rules. 

The third difference is that INLA bases inferences off of a modified model, with noisy parameter $\parambigS = (\addpredS,\parambig)$. Evaluating the Gaussian approximation $\gaussapprox(\parambigS|\data,\paramsmall)$ required to compute (\ref{eqn:gaussapprox}), (\ref{eqn:laplace}), and (\ref{eqn:nested}) under the modification employed by INLA therefore requires computing, storing, and decomposing the matrix $\lphessWS(\paramsmall) = \uprecS(\paramsmallprior) + \llhessWS(\WmodeS_{\paramsmall},\paramsmalllik)\in\R^{(n+\Wdim)\times(n+\Wdim)}$, where
\*[
\uprecS(\paramsmallprior) = \tau\begin{pmatrix}\I_{n} & -\Zmat \\ -\Zmat\tpose & \tau^{-1}\uprec(\paramsmallprior) + \Zmat\tpose\Zmat\end{pmatrix}, \qquad
\llhessWS(\WmodeS_{\paramsmall},\paramsmalllik) = \begin{pmatrix} \llhesseta(\addpredSmode_{\paramsmall},\paramsmalllik) & \zero \\ \zero & \zero \end{pmatrix}.
\]
Here $\parambigS\sim\Normal\left\{(\addpred,\zero),\uprecS(\paramsmallprior)^{-1}\right\}$, $\llhesseta(\addpred,\paramsmalllik)=-\partial^{2}_{\addpred}\log\pi(\data|\addpred,\paramsmalllik)\in\R^{n\times n}$, and hence $\llhessWS(\WmodeS_{\paramsmall},\paramsmalllik)$ depends on $\WmodeS_{\paramsmall}$ through $\addpredSmode_{\paramsmall}$. Crucially, in an LGM, the matrix $\llhesseta(\addpredSmode_{\paramsmall},\paramsmalllik)$ is diagonal, owing to the conditional independence of $\datai_{i}$ given $\eta_{i}$. The feasibility of computations under the modified model therefore depends on the conditional independence constraint enforced in an LGM.

In ELGMs, however, this constraint is relaxed, and as a result, the $n\times n$ matrix $\llhesseta(\addpredSmode_{\paramsmall},\paramsmalllik)$ is no longer diagonal, and in fact may be dense. To summarize, basing inferences off of the modified model has several implications:
\begin{enumerate}
	\item It becomes challenging to fit models with more complex dependence structures than LGMs allow, because of the need to store and manipulate large, potentially dense matrices,
	\item It becomes challenging to fit LGMs to large datasets, because the size of the sparse matrices involved scales linearly with $n$ for every such model fit,
	\item It becomes challenging to study the theoretical properties of the approximations, because convergence results pertaining to the marginal Laplace approximation \citep{tierney} do not apply in this context \citep{higherorderlaplace}.
\end{enumerate}

We now define the class of ELGMs and describe our approach to inference for them which mitigates these challenges.


\section{Extended Latent Gaussian Models}\label{sec:elgms}

\subsection{Definition of ELGMs}\label{subsec:elgms:definition}

In ELGMs, the vector of additive predictors is redefined as $\addpred = (\addpredi_{j})_{j\in[\etadim_{n}]}$, with dimension $\etadim_{n}\in\N$ depending on, but not necessarily equal to, $n$. This dependence is unrestricted, and we may have in any particular example $\etadim_{n}<n,\etadim_{n}>n$, or $\etadim_{n}=n$. Each mean response $\mu_{i}$ will now depend on some \emph{subset}, $\J_{i}\subseteq[\etadim_{n}]$, of indices of $\addpred$. We require that $\bigcup_{i=1}^{n}\J_{i}=[\etadim_{n}]$ and $1\leq\abs{\J_{i}}\leq\etadim_{n}$, but otherwise enforce no restrictions on the $\J_{i}$. This means, informally, that observation means may depend on multiple, shared additive predictors, in a very general manner. In contrast, LGMs always have $\J_{i}=\bracevec{i},i\in[n]$, and $\etadim_{n}=n$. The inverse link function $g(\cdot)$ is redefined for each observation to be a many-to-one mapping $g_{i}:\R^{\abs{\J_{i}}}\to\R$, such that $\mu_{i} = g_{i}(\addpred_{\J_{i}})$.

Extended Latent Gaussian Models take the following form:
\begin{equation}\begin{aligned}\label{eqn:elgm}
\datai_{i} | \addpred_{\J_{i}},\paramsmalllik &\overset{ind}{\sim}\pi(\datai_{i}|\addpred_{\J_{i}},\paramsmalllik), i\in[n], \ \mu_{i} = \EE(\datai_{i} | \addpred_{\J_{i}}) = g_{i}(\addpred_{\J_{i}}), \ \addpredi_{j} = \covx_{j}\tpose\mb{\beta} + \sum_{q=1}^{r}u_{q}(z_{jq}), j\in[\etadim_{n}],
\end{aligned}\end{equation}
which may be contrasted with the LGM (\ref{eqn:lgm}).

The potentially complex dependence between $\mu_{i}$ and $\addpred$ introduces computational burden depending directly on the sets $\J_{i}$. Specifically, the sparsity pattern of $\llhesseta(\addpred,\paramsmalllik)$ is determined as follows: for any $k,l\in[\etadim_{n}],$ $$\exists i\in[n]: k,l\in\J_{i}\implies\llhesseta(\addpred,\paramsmalllik)_{kl} \neq 0.$$ Because of this relationship, the $\etadim_{n}$-dimensional $\llhesseta(\addpred,\paramsmalllik)$ matrix becomes denser as the model becomes more complex, as specifically measured by the complexity of the index sets $\J_{i}$. This in turn affects the computations required for approximate Bayesian inference in ELGMs.

\subsection{Computational Considerations}\label{subsec:elgms:computational}

Algorithm \ref{alg:implementation} shows the full approximation procedure for ELGMs. We highlight two important points here. First, we base inferences off of $\parambig$ instead of $\parambigS$, and hence the sparsity structure of the Hessian $\lphessW(\paramsmall) = -\partial^{2}_{\parambig}\log\pi(\Wmode_{\paramsmall}|\data,\paramsmall) = \uprec(\paramsmallprior) + \llhessW(\Wmode_{\paramsmall},\paramsmalllik)$ is determined both by that of $\uprec(\paramsmallprior)$ and
\*[
\llhessW(\parambig,\paramsmalllik) = \Zmat\tpose\llhesseta(\addpred,\paramsmalllik)\Zmat,
\]
which is (now) of dimension $m\times m$ where $m = \text{dim}(\parambig)$, for any $n$. While $\llhessW(\parambig,\paramsmalllik)$ may be less sparse than $\llhessWS(\parambigS,\paramsmalllik)$ in general, and hence less amenable to the use of efficient sparse matrix algorithms \citep{fastsamplinggmrf}, it is also much smaller, with dimension $n$ less than that of $\llhessWS(\parambigS,\paramsmalllik)$. This has the following implications:
\begin{enumerate}
	\item It becomes feasible to fit ELGMs, because the dimension of the large, potentially dense matrices involved in fitting is reduced by $n$ compared to if inferences were based on $\parambigS$,
	\item It becomes more efficient to fit LGMs to larger datasets (Table \ref{tab:inlacompare} in \S\ref{sec:comp}), for this same reason,
	\item The theoretical properties of the approximation error are more feasible to study (\S\ref{sec:theory}) because the convergence of the marginal Laplace approximation is established \citep{tierney}.
\end{enumerate} 

We still exploit sparsity when available on a problem-specific basis, using the efficient routines in the \texttt{Matrix} package for \texttt{R} \citep{matrix}. Further, the reduced size of matrices involved is expected to improve computational efficiency of our approximations at larger sample sizes, where this difference is more pronounced. We investigate these notions empirically in \S\ref{sec:comp}.

Second, we choose Gauss-Hermite quadrature for constructing the rule $\quadpointset(\paramsmalldim,\quadnum),\weight_{\quadnum}(\cdot)$ that is adapted to the renormalization of $\laplaceapprox(\paramsmall,\data)$. A number of suitable extensions $\quadpointset(\paramsmalldim,\quadnum)$ of the univariate rule $\quadpointset(1,\quadnum)$ to $\paramsmalldim$ dimensions are permitted, including product \citep{adaptive}, sparse \citep{heiss08sparse}, and nested \citep{nested} rules. Beyond being readily available in standard software \citep{mvquad}, the asymptotic properties of the adaptive GHQ rule were recently established by \cite{aghq}, and we use these results in the proof of Theorem \ref{thm:convergence} (\S\ref{sec:theory}; Online Supplement A).

\subsection{Implementation}\label{subsec:elgm:implementation}

Algorithm \ref{alg:implementation} is implemented in the \texttt{aghq} package for the \texttt{R} language. The core of the implementation requires computing $\laplacemode$ and $\laplacehess(\laplacemode)$, which requires optimization and two numerical derivatives of $\laplaceapprox(\paramsmall,\data)$. This is complicated by the fact that each evaluation of $\laplaceapprox(\paramsmall,\data)$ requires evaluating $\gaussapprox(\parambig|\data,\paramsmall)$, which requires further optimization and derivatives of $\pi(\parambig,\paramsmall,\data)$ with respect to $\parambig$. 

Our recommendation is to implement $-\log\pi(\parambig,\paramsmall,\data)$ using the \texttt{TMB} package for automatic differentiation in \texttt{R} \citep{tmb}, which provides two automatic derivatives of $\log\pi(\parambig,\paramsmall,\data)$, automatic evaluation of $\laplaceapprox(\paramsmall,\data)$, and---critically---an automatic derivative of $\laplaceapprox(\paramsmall,\data)$ that avoids repeatedly computing $\Wmode_{\paramsmall}$. When a \texttt{TMB} template is not available, we use trust region optimization to compute $\Wmode_{\paramsmall}$, with sparse \citep{trustoptim} or dense \citep{trustdense} matrix algebra depending on the application, and numerical derivatives of $\laplaceapprox(\paramsmall,\data)$. Assumption 2 in Online Supplement A states that the objective function $\pi(\parambig,\paramsmall,\data)$ is log-concave (locally, with high probability as $n\to\infty$), making trust region optimization stable and efficient for this application.

{\spacingset{1}
\begin{algorithm}[p]
\textbf{Input}:
\begin{itemize}
\item[] Likelihood $\pi(\data|\parambig,\paramsmall)$, precision matrix $\uprec(\paramsmallprior)\in\R^{\Wdim\times\Wdim}$, prior $\pi(\paramsmall)$, Hessian $\lphessWraw(\paramsmall) = -\partial^{2}_{\parambig}\log\pi(\parambig,\paramsmall,\data)$, number of samples $B\in\N$, number of quadrature points $\quadnum\in\N$ and base quadrature rule $\quadpointset(\paramsmalldim,\quadnum)$.
\end{itemize}
\textbf{Compute}:
\begin{enumerate}
	\item $\laplacemode = \text{argmax}_{\paramsmall} \ \laplaceapprox(\paramsmall,\data)$ (Equation \ref{eqn:laplace}) using numerical optimization. For each evaluation of $\laplaceapprox(\paramsmall,\data)$ at point $\paramsmall$, \textbf{compute}:
	\begin{enumerate}
		\item $\Wmode_{\paramsmall} = \text{argmax}_{\parambig} \ \pi(\parambig,\paramsmall,\data)$ using trust region optimization (\S\ref{subsec:elgm:implementation}),
		\item Hessian $\lphessW(\paramsmall)$ analytically or using automatic differentiation,
		\item $\gaussapprox(\Wmode_{\paramsmall}|\data,\paramsmall) = (2\pi)^{-\Wdim/2}\abs{\lphessW(\paramsmall)}^{1/2}$, and hence $\laplaceapprox(\paramsmall,\data) = \pi(\Wmode_{\paramsmall},\paramsmall,\data)/\gaussapprox(\Wmode_{\paramsmall}|\data,\paramsmall)$.
	\end{enumerate}
	\item $\laplacehess(\paramsmall) = -\partial^{2}_{\paramsmall} \log\laplaceapprox(\paramsmall,\data)$ using numeric or automatic differentiation,
	\item The lower Cholesky triangle $\laplacechol$ of $\laplacehess(\laplacemode)^{-1} = \laplacechol\laplacechol\tpose$, directly.
\end{enumerate}
\textbf{Output}:
\begin{enumerate}
	\item Posterior of summaries of $\laplaceapprox(\paramsmall|\data)$, using further numerical integration,
	\item Independent samples $(\parambig_{b})_{b\in[B]}\overset{iid}{\sim}\approxpi(\parambig|\data)$ (Equation \ref{eqn:nested}):
\item[] \textbf{Input}: $B\in\N$,  $\lambda(\quadpoint) = \abs{\laplacechol}\laplaceapprox(\laplacechol\quadpoint+\laplacemode|\data)\weight_{\quadnum}(\quadpoint), \quadpoint\in\quadpointset(\paramsmalldim,\quadnum)$.
\item[] \textbf{For}: $b = 1,\ldots,B$:
	\begin{enumerate}
		\item[a.] Draw $\quadpoint_{b}$ from $\quadpointset(\paramsmalldim,\quadnum)$ with probabilities $\PP(\quadpoint_{b} = \quadpoint) = \lambda(\quadpoint), \quadpoint\in\quadpointset(\paramsmalldim,\quadnum)$,
		\item[b.] Draw $\parambig_{b}$ from $\gaussapprox(\cdot|\data,\laplacechol\quadpoint_{b}+\laplacemode)$ \citep{fastsamplinggmrf},
	\end{enumerate}
\end{enumerate}
\caption{Fitting Extended Latent Gaussian Models}
\label{alg:implementation}
\end{algorithm}
}


\section{Theoretical Considerations}\label{sec:theory}

\subsection{Approximation error}

Inferences made from Algorithm \ref{alg:implementation} are based on the approximations $\laplaceapprox(\paramsmall|\data)$ and $\approxpi(\parambig|\data)$. Convergence results pertaining to $\laplaceapprox(\paramsmall|\data)$ are standard, see \citet{tierney,validitylaplace,higherorderlaplace,laplaceerror,approximatelikelihood} as well as Equations (3) and (5) in Online Supplement A, but note that these results do not immediately apply when inferences are based on $\parambigS$ instead of $\parambig$, since $\text{dim}(\parambigS) = O(n)$ \citep{higherorderlaplace,laplaceerror}. Even when basing inferences off of $\parambig$ as we do, convergence of $\approxpi(\parambig|\data)$ is non-trivial. 

Define the total variation error $\tvnorm{\pi(\parambig|\data) - \approxpi(\parambig|\data)} = \sup_{\borelset\in\borelsets{\Wdim}}\abs{\int_{\borelset}\pi(\parambig|\data) - \approxpi(\parambig|\data)d\parambig}$, where $\borelsets{\Wdim}$ are the Borel sets on $\R^{\Wdim}$. We have the following result:
\begin{theorem}\label{thm:convergence}
Under Assumptions 1 -- 7 (Online Supplement A), $\tvnorm{\pi(\parambig|\data) - \approxpi(\parambig|\data)} = o_{p}(1).$
\end{theorem}
The proof of Theorem \ref{thm:convergence} is given in Online Supplement A. Practically, Theorem \ref{thm:convergence} guarantees convergence in probability of the approximate coverage $\widetilde{\alpha} = \int_{\borelset}\approxpi(\parambig|\data)d\parambig$ of any credible set $\borelset\in\borelsets{\Wdim}$ to its nominal level $\alpha = \int_{\borelset}\pi(\parambig|\data)d\parambig$.


Our nested approximation strategy (Equation \ref{eqn:nested}) assumes that each individual Gaussian (\ref{eqn:gaussapprox}) and Laplace (\ref{eqn:laplace}) approximation converges in probability to their exact counterpart, which is the case when the likelihood satisfies, respectively, Assumption 6 and Assumptions 1--5, 7 in Online Supplement A. In that case, Theorem \ref{thm:convergence} guarantees that our nested approximation (\ref{eqn:nested}) preserves this property, showing the convergence to zero in probability of the total variation measure between the approximate and exact posterior. This is a non-trivial result, for two reasons, both of which are also present in the method of, but not discussed by, \citet[\S4.1]{inla}. 

First, the Gaussian approximation $\gaussapprox(\parambig|\data,\paramsmall)$ is evaluated at $\paramsmall = \laplacechol\quadpoint+\laplacemode,\quadpoint\in\quadpointset(\paramsmalldim,\quadnum)$, which depend on $\data$, and hence is an approximation to the posterior density $\pi(\parambig|\data,\laplacechol\quadpoint+\laplacemode)$ based on the likelihood $\pi(\data|\parambig,\laplacechol\quadpoint+\laplacemode)$. Even under the strong assumption (which we do not make, see Assumptions 1 -- 7 in Online Supplement A) that the model is well-specified, in the sense that $\data$ has joint density $\pi(\data|\parambigtrue,\paramsmalltrue)$ for some $\parambigtrue,\paramsmalltrue$, this likelihood is misspecified, because it is not evaluated at $\paramsmalltrue$, and standard Bernstein von-Mises theory \citep[Ch. 10.2]{vandervaart} does not immediately apply. This notion also applies to the Gaussian and Laplace approximations employed in this manner by \citet{inla} and \citet{simplifiedinla}, although this is not discussed.

Second, the quadrature points and weights are adapted to integrate $\laplaceapprox(\paramsmall,\data)$, but are then used a second time to integrate $\gaussapprox(\parambig|\data,\paramsmall)\laplaceapprox(\paramsmall|\data)$ (for any $\parambig)$. This strategy was initially proposed by \citet{inla} and has obvious computational benefits, but requires theoretical justification. To our knowledge, the theoretical implications of this have not been investigated in any other work. In proving Theorem \ref{thm:convergence}, we explicitly handle this theoretical complexity.

\subsection{Limitations and practical considerations}\label{subsec:limitations}

Beyond the challenges present in establishing Theorem \ref{thm:convergence}, as noted by \citet[\S4.1]{inla}, there are a number of practical issues with many applications of LGMs that make study of the approximation error challenging. Similar issues also apply to ELGMs, and here we briefly expand on this discussion. The two main challenges are:
\begin{enumerate}
	\item Despite basing inferences on $\parambig$ instead of $\parambigS$, many specific ELGMs have $\text{dim}(\parambig)$ increasing with $n$ by construction, and the convergence of the Laplace approximation is not immediately implied by standard results \citep{higherorderlaplace},
	\item Even under the assumption that the model (\ref{eqn:elgm}) is well-specified, the basis function expansion $u_{q}(z) = \sum_{j=1}^{d_{q}}\basisfunction_{qj}(z)\uui_{qj}$ induces misspecification into the model, and model-specific work is required to validate Assumptions 1 -- 7.
\end{enumerate}
Theorem \ref{thm:convergence} can be expected to hold for well-specified models with fixed parameter dimension. However, it is not the case that Theorem \ref{thm:convergence} is false for models exhibiting one or both of the above challenges. Rather, efforts must be made on a model-specific basis to establish whether the assumptions underlying Theorem \ref{thm:convergence} hold. The need to verify complex analytical criteria on a model-specific basis in order to apply general convergence thoery is not unprecedented; for example, \citet{approximatelikelihood,misspec}. Asymptotics in spatial problems are especially challenging \citep{maternbad,glmmnormal,lgcpconvergence,pcmatern}.

Model-specific challenges are not addressed by Theorem \ref{thm:convergence}. We emphasize that (a) Theorem \ref{thm:convergence} reduces the task of establishing posterior convergence to that of verifying whether the three \emph{individual} approximations converge for a given model, by addressing all the complexities associated with applying these approximations in a non-standard manner, and (b) the problems with nested approximations that Theorem \ref{thm:convergence} addresses are all present when fitting LGMs using INLA, and remain unaddressed in that context.

Some results exist that are relevant to the examples given in \S\ref{sec:comp} and \S\ref{sec:examples}. \citet{approximatelikelihood} shows that Assumption 7 holds for a Bernoulli Generalized Linear Mixed Model of the type we consider in \S\ref{sec:comp}, and reinforces the claim that the model-specific calculations required are often non-trivial. \citet{laplaceerror} expands on the general application of Laplace approximations in models where $\text{dim}(\parambig)$ increases with $n$, which is also relevant to the examples of \S\ref{subsec:coxph} and \S\ref{subsec:astro}. \citet{spde} analyze the error in the basis function approximation used in \S\ref{subsec:aggspatial}, and \citet{lgcpconvergence} discuss convergence of the approximate likelihood used in that example.

\section{Computational Comparisons}\label{sec:comp}

In this section we compare our procedure (Algorithm \ref{alg:implementation}) to the \RINLA{} implementation of the INLA method of \citet{inla} for a model which both can fit, and compare both to \MCMC{} using \rstan{} \citep{stan}. We find that our procedure, which does not make inferences based on the modified model (\S\ref{subsec:intro:computational}, \S\ref{subsec:prelim:approximateinference}), runs faster than \RINLA{} (Table \ref{tab:inlacompare}), and that the difference becomes more pronounced at larger sample sizes.

We use the testing version of \RINLA{} accessed on 2021/10/22 with the \texttt{PARDISO} sparse matrix library. Computations were done on a private virtual server with 180 Gb of memory and 16 cores. We parallelize our likelihood computations across $16$ cores. \RINLA{} and \texttt{PARDISO} were run in parallel using the \texttt{num.threads} and \texttt{num.blas.threads} options using $16$ cores as well, except when $n>1,000,000$ when a single core was used to avoid crashing. We set \texttt{strategy='gaussian'} in \RINLA{}, yielding an approximation most similar to $\approxpi(\parambig|\data)$ out of the available approximations, and hence the fairest comparison of computation time between the two approaches. We implement \MCMC{} using the self-tuning No-U-Turn sampler under the default control parameters \citep{nuts} in the \rstan{} language \citep{stan}, through the \texttt{tmbstan} package \citep{tmbstan}, using the same \texttt{TMB} template as we use to implement Algorithm \ref{alg:implementation}. 

We perform our comparisons by fitting a Bernoulli Generalized Linear Mixed Model to a large dataset. A complete record of every person who was discharged from a publicly funded drug or alcohol treatment facility in the United States between 2006 -- 2011 and was listed as having used opioids was obtained from the Inter-University Consortium for Political and Social Research through the University of Michigan \citep{opioiddata}. The dataset contains person-level records of the year of admission, reason for discharge (including successful completion of the program), and gender, race, and living arrangement (homeless, independent or dependent), for each such person. There are $n = 7,283,575$ records after filtering out missing data. We quantify the association between the probability of a subject successfully completing their program and their gender, race, and living arrangements, using discrete, nested random effects to account for correlation between outcomes from subjects living in the same state and the same town. 

Denote by $\datai_{i}$ the observed indicator of whether the $i^{th}$ subject completed their treatment, $i = 1\ldots n$, and the vector of covariates for each subject by $\covx_{i}$. There are $d_{1} = 47$ states and $d_{2} = 262$ towns with reported data, and we denote by $z_{i1}\in[d_{1}]$ the state and by $z_{i2}\in[d_{2}]$ the town in which the $i^{th}$ subject lives. We consider the following model:
\begin{equation}\begin{aligned}
Y_{i}|\eta_{i} &\overset{ind}{\sim}\text{Bernoulli}(p_{i}), i\in[n], \\
\log\left(\frac{p_{i}}{1 - p_{i}}\right) &= \eta_{i} = \mb{x}_{i}^{T}\mb{\beta} + u_{1}(z_{i1}) + u_{2}(z_{i2}), \\
\uui_{1j}|\sigma_{1} &\iid \text{N}(0,\sigma_{1}^{2}), j\in[d_{1}], \\
\uui_{2t}|\sigma_{2} &\iid \text{N}(0,\sigma_{2}^{2}), t\in[d_{2}], \\
\end{aligned}\end{equation}
where $u_{1}(z_{i1}) = \uui_{1j}$ whenever $z_{i1} = j,j\in[d_{1}]$, and $u_{2}(z_{i2}) = \uui_{2t}$ whenever $z_{i2} = t,t\in[d_{2}]$, so that $\uu_{1} = (\uui_{1j})_{j\in[d_{1}]},\uu_{2} = (\uui_{1t})_{t\in[d_{2}]}$, and $\uu = (\uu_{1},\uu_{2})\in\R^{d}$ where $d = d_{1}+d_{2}$. The regression coefficients all correspond to categorical predictors, with a total of $\covxdim = 7 + 1$ non-reference levels and an intercept. The index sets are $\J_{i} = \bracevec{i},i\in[n]$ and hence this model is a LGM, compatible with the methodology of \citet{inla}. The full parameter vector is $\parambig = (\uu,\mb{\beta})$, and $\text{dim}(\parambig) = d_{1}+d_{2}+p = 47 + 262 +8= 317$. In contrast, when running \RINLA{}, inferences are made based on the modified model $\parambigS = (\addpredS,\parambig)$, having $\text{dim}(\parambigS) = 317 + n = 7,283,892$ for the full dataset, resulting in much larger matrices to be stored and decomposed during fitting.

We fit the model to subsamples of the dataset ranging from a size of $n=1,000$ to $n=1,000,000$. Ten different subsamples were taken of each size, and the computing times and posterior distributions were compared. The full dataset could not be fit with \rstan{}, and caused \RINLA{} to crash except when restricted to a single core. Table \ref{tab:inlacompare} shows the mean and standard deviation of computing times for each scenario. The \ELGM{} procedure (Algorithm \ref{alg:implementation}) is faster than \RINLA{}, with the difference increasing as the size of the dataset increases. For $n=1,000,000$, the \ELGM{} is roughly six times faster, and the full dataset with $n=7,283,575$ is $15$ times faster with \ELGM{}, with the caveat that \RINLA{} ran in single core mode.
 
Table \ref{tab:kstable} compares the similarity of the posterior distributions for the three inference methods, as measured by the KS statistic (the maximum distance between empirical CDFs). The \ELGM{} is, for the most part, closer to \rstan{} than \RINLA{} for the $\sigma$ parameters, and the posterior for the $\beta$ parameters are broadly comparable. For $n=1,000,000$ observations, \ELGM{} is perceptible closer to \rstan{} than \RINLA{} for the $\sigma$ parameters, and the discrepancy between \RINLA{} and \ELGM{} is substantial when run on the full dataset. The posterior for $\sigma_1$ appears to be over-estimated by \RINLA{} when using the full dataset, as the posterior median of $3.19$ and $95\%$ credible interval of $(2.86, 3.46)$ is well above the \RINLA{} estimate from $1,000,000$ observations of $2.14$ $(1.35, 3.29)$, the \ELGM{} estimates of $1.88$ $(1.17, 2.88)$ and $2.14$ $(1.33, 3.29)$ for the full and reduced datasets respectively, and \rstan{}'s estimates from the reduced dataset of $2.14$ $(1.35, 3.30)$.

\begin{table}[p]
\centering
\begin{tabular}{|l|ll|ll|}
\hline
    & \multicolumn{2}{c|}{Time, seconds} & \multicolumn{2}{c|}{Iterations} \\      
$n$ & \ELGM{} & \RINLA{} & \ELGM{} & \RINLA{} \\
\hline
1,000 & 1.25 (0.235) & 2.57 (0.230) & 308 & 630 \\
10,000 & 2.17 (0.193) & 4.75 (0.157) & 176 & 385 \\
100,000 & 9.44 (0.645) & 27.7 (2.47) & 36.3 & 107 \\
1,000,000 & 100 (4.75) & 641 (18.5) & 10.2 & 65.3 \\
7,283,575 & 926 (5.49) & 14,169 & -- & -- \\ 
\hline
\end{tabular}
\caption{Mean (std. dev) run times in seconds for $10$ runs each of \ELGM{} and \RINLA{}, run in parallel using $16$ cores, on subsets of the opioid treatment centre data of \S\ref{sec:comp}. For $n=7,283,575$, \RINLA{} was run once, single-threaded (to avoid crashing), and \rstan{} was not run, to avoid excessive computation times. Also shown are the number of iterations of \rstan{} that could have been run in the same time it took to fit \ELGM{} and \RINLA{}, based on a single run of $1,000$ iterations of \rstan{} at each sample size.}
\label{tab:inlacompare}
\end{table}

\begin{table}[p]
\centering
\hspace*{-.5in}
\begin{tabular}{|ll|rrrrrrrr|rr|}
\hline
$n$ & Methods & $\beta_{0}$ & $\beta_{1}$ & $\beta_{2}$ & $\beta_{3}$ & $\beta_{4}$ & $\beta_{5}$ & $\beta_{6}$ & $\beta_{7}$ & $\sigma_{1}$ & $\sigma_{2}$ \\
\hline
\multirow{3}*{$1,000$} & \ELGM{}/\rstan{} & 0.036 & 0.011 & 0.021 & 0.026 & 0.022 & 0.048 & 0.033 & 0.102 & 0.046 & 0.053\\
 & \RINLA{}/\rstan{} & 0.067 & 0.012 & 0.043 & 0.015 & 0.036 & 0.014 & 0.028 & 0.063 & 0.051 & 0.093\\
 & \ELGM{}/\RINLA{} & 0.099 & 0.017 & 0.050 & 0.021 & 0.021 & 0.058 & 0.056 & 0.048 & 0.019 & 0.047\\
\hline
\multirow{3}*{$10,000$} & \ELGM{}/\rstan{} & 0.019 & 0.016 & 0.031 & 0.035 & 0.009 & 0.040 & 0.023 & 0.031 & 0.017 & 0.069\\
 & \RINLA{}/\rstan{} & 0.020 & 0.022 & 0.010 & 0.012 & 0.020 & 0.009 & 0.017 & 0.031 & 0.011 & 0.101\\
 & \ELGM{}/\RINLA{} & 0.027 & 0.018 & 0.031 & 0.030 & 0.023 & 0.046 & 0.012 & 0.056 & 0.012 & 0.043\\
\hline
\multirow{3}*{$100,000$} & \ELGM{}/\rstan{} & 0.049 & 0.013 & 0.023 & 0.007 & 0.025 & 0.024 & 0.012 & 0.028 & 0.013 & 0.016\\
 & \RINLA{}/\rstan{} & 0.026 & 0.024 & 0.017 & 0.011 & 0.018 & 0.014 & 0.023 & 0.017 & 0.013 & 0.031\\
 & \ELGM{}/\RINLA{} & 0.037 & 0.021 & 0.014 & 0.013 & 0.025 & 0.019 & 0.016 & 0.031 & 0.015 & 0.031\\
\hline
\multirow{3}*{$1,000,000$} & \ELGM{}/\rstan{} & 0.017 & 0.009 & 0.016 & 0.009 & 0.011 & 0.010 & 0.010 & 0.017 & 0.018 & 0.009\\
 & \RINLA{}/\rstan{} & 0.040 & 0.011 & 0.012 & 0.012 & 0.009 & 0.008 & 0.013 & 0.010 & 0.016 & 0.035\\
 & \ELGM{}/\RINLA{} & 0.041 & 0.009 & 0.010 & 0.008 & 0.015 & 0.012 & 0.014 & 0.025 & 0.030 & 0.038\\
\hline
$7,283,575$ & \ELGM{}/\RINLA{} & 0.329 & 0.019 & 0.016 & 0.012 & 0.014 & 0.022 & 0.011 & 0.011 & 0.957 & 0.329\\
\hline
\end{tabular}
\hspace*{-.5in}
\caption{KS statistic (maximum distance between empirical CDFs) for \ELGM{}, \RINLA{}, and \rstan{}, for the ten marginal posterior distributions at each sample size, for the Opioid example of \S\ref{sec:comp}. The \ELGM{} results are generally as close to \rstan{} as \RINLA{}, but computed in a fraction of the time (Table \ref{tab:inlacompare}).}
\label{tab:kstable}
\end{table}


\section{Examples of Extended Latent Gaussian Models}\label{sec:examples}

In this section we demonstrate the breadth of the ELGM class and practical utility of our approximate inference method for it through three challenging examples. We fit a point process model to spatially-aggregated data \citep{lgcpagg,disaggregation} and compare the accuracy of our approach to MCMC. We fit a Cox Proportional Hazards model with partial likelihood for mapping spatial variation in Leukaemia survival times, which is an example of a model in which the Hessian of the log-likelihood is fully dense and hence the most computationally-intensive type of ELGM. We then fit the Galactic Mass Estimator model of \citet{gwen3} and \citet{gwen4} for estimating the mass of the Milky Way galaxy in the presence of multivariate measurement uncertainties, a challenging model. All three examples are beyond the class of LGMs but belong to the class of ELGMs. The computations make use of the \texttt{aghq} package in the \texttt{R} language, and code for all examples is available from \url{https://github.com/awstringer1/elgm-paper-code}.


\subsection{Spatially Aggregated Point Process Data}\label{subsec:aggspatial}

A spatial point process is a stochastic process which generates random points within a fixed study area. Often, for reasons of privacy or lack of available resources, the exact point locations are not recorded and are instead aggregated to counts within predefined  regions within this area. In such cases, covariate information may be available at a different spatial resolution than the response, complicating inference. Spatial \emph{downscaling} or \emph{disaggregation} is a challenging task where an aggregated response is combined with fine-scale covariate information to infer spatial variation in a phenomenon of interest at a higher resolution than the observed counts. This task has been approached using frequentist methods \citep{aggspatial,rootgaussian} and Bayesian inference using specialized MCMC algorithms \citep{lgcpagg,geostataggdata}. 

More recently, \citet{disaggregation} develop software for approximate Bayesian inference in these models where a single Laplace approximation (corresponding to adaptive Gauss-Hermite quadrature with $k=1$ point) is used for the joint posterior of all model parameters, making use of the \texttt{TMB} package \citep{tmb}. They describe an example of inferring Malaria incidence in Madagascar using aggregated case counts and fine-scale environmental covariates. We fit this example here within the ELGM framework, using more than one point for the adaptive quadrature, and obtaining more accurate approximations to the posteriors when compared to a long MCMC run. 

Denote the study region of Madagascar by $\M\subset\R^{2}$ and define an inhomogenous Poisson point process $X = \bracevec{X(\loc):\loc\in\M}$ with rate function $\lambda(\cdot)$. We do not observe point locations $X(\loc)$, and instead observe case counts $\data$ with $\datai_{i} = \norm{\loc\in\M: X(\loc)\in S_{i}}$ for predefined regions $S_{i}\subset\M$ with $\cup_{i=1}^{n}S_{i} = \M$ and $S_{i}\cap S_{j} = \emptyset$ for $i\neq j\in[n]$. For any $\loc\in\M$ define spatially-varying covariates $\covx(\loc): \R^{2}\to\R^{p}$ corresponding to elevation, vegetation index and land surface temperature, with $p = 4$ including the intercept, and known population offset function $P(\loc): \R^{2}\to\R$ such that the population in any fixed area $S\subseteq\M$ is $\int_{S}P(\loc)d\loc$. We wish to infer incidence risk $\lambda(\cdot)$ using the spatially-aggregated log-Gaussian Cox process:
\*[
Y_{i}|\lambda(\cdot),v_{i}&\ind\text{Poisson}\left[ \exp(v_{i})\int_{S_{i}}P(\loc)\lambda(\loc)d\loc\right], \ \log\lambda(\loc) = \beta_{0} + \mb{x}(\loc)\tpose\mb{\beta} + u(\loc), \\
u(\loc) &\sim \mathcal{GP}\left\{0,\text{M}_{\nu}(\cdot;\sigma,\rho)\right\}, \ \text{Cov}\{u(\loc + \mb{h}),u(\loc)\} = \Matern_{\nu}(\norm{\mb{h}};\sigma,\rho), \loc\in\M,\mb{h}\in\M, \\
v_{i} &\overset{iid}{\sim}\Normal(0,\tau^{-1}), \ \tau^{-1/2}\sim \text{Exp}(\xi_{\tau}), \sigma\sim \text{Exp}(\xi_{\sigma}), \rho^{-1}\sim \text{Exp}(\xi_{\rho}).
\]
Here $\Matern_{\nu}(\cdot;\sigma,\rho)$ is a Matern covariance function with fixed shape parameter $\nu = 1$ in the parametrization of \citet{diseasemapping}. The polygon-level effects $v_{i}$ are included to account for overdispersion in the polygon counts.

The risk surface $\lambda(\mb{s})$ must be discretely inferred, and this requires approximations to the continuously-defined covariates $\mb{x}(\mb{s})$ and spatial process $u(\mb{s})$. For the spatial process we use a basis-function representation \citep{spde}:
\begin{equation}\label{eqn:basisfunction}
u(\loc) \approx \sum_{j=1}^{d}\basisfunction_{j}(\loc)\uui_{j},
\end{equation}
where $\basisfunction_{j}(\cdot)$ are fixed, known piecewise-linear functions. The unknown parameter is $\uu = (\uui_{j})_{j\in[d]}$. This approximation is constructed on a \emph{mesh} with triangular cells $\M_{j},j\in[d]$ such that $\M\subseteq\cup_{j=1}^{d}\M_{j}$ and $\M_{j}\cap\M_{k} = \emptyset, j\neq k\in[d]$ \citep{inlasoftware}. See \citet{spde} and \citet{inlasoftware} for further details.

\begin{figure}[p]
	\centering
	\begin{subfigure}[t]{0.3\textwidth}
		\centering
		\includegraphics[height=2in]{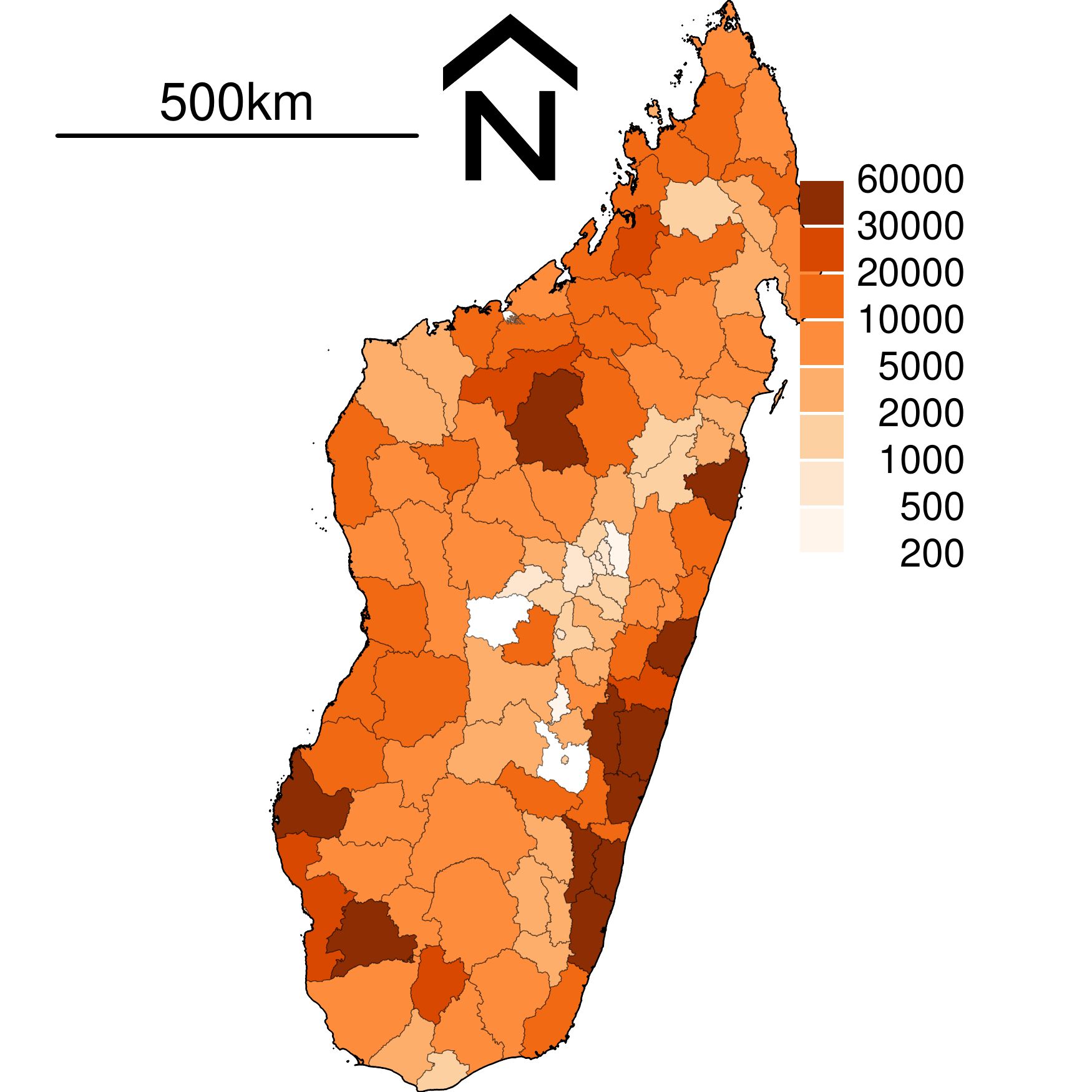}
		\caption{Malaria case counts $Y_{i}$}
	\end{subfigure}
	\begin{subfigure}[t]{0.3\textwidth}
		\centering
		\includegraphics[height=2in]{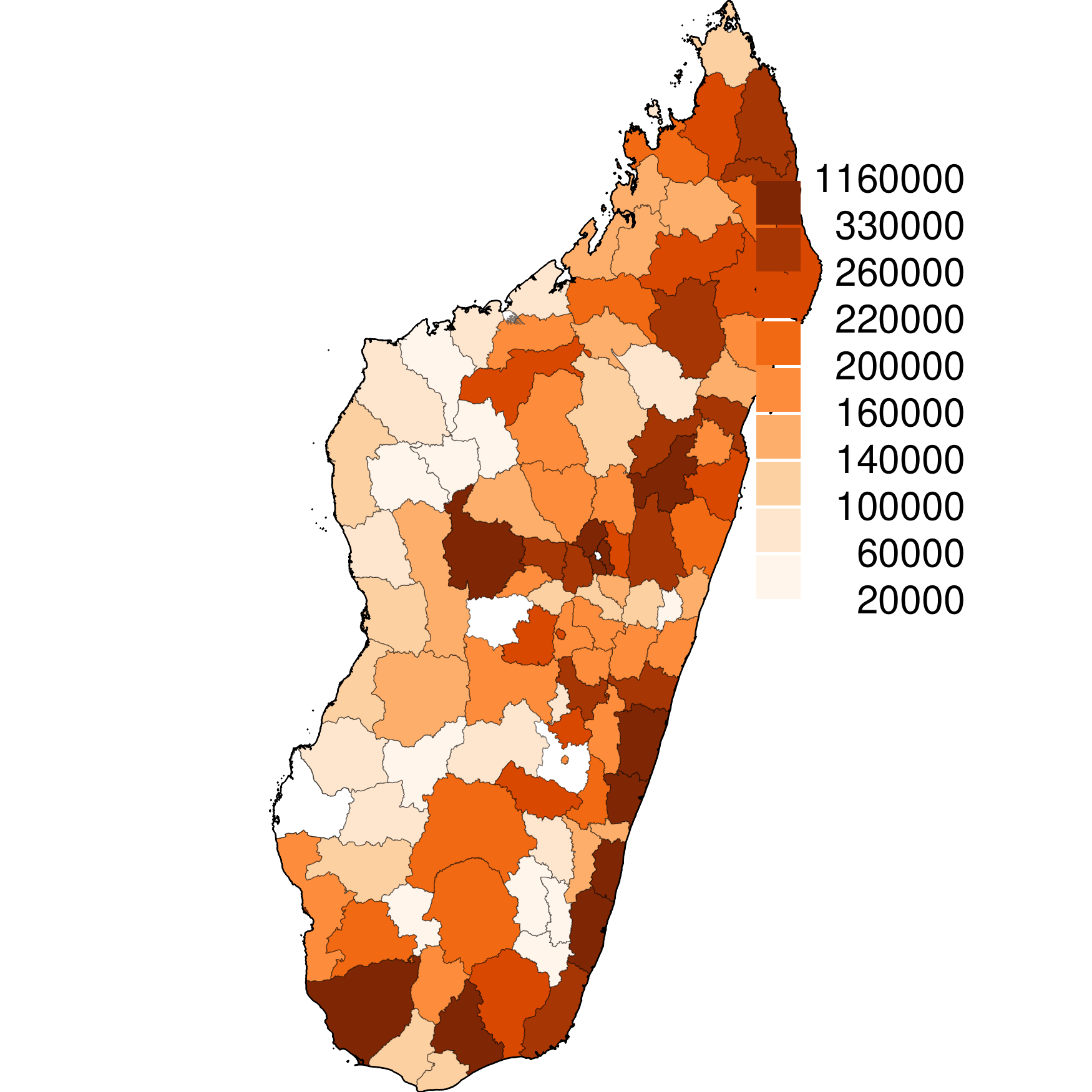}
		\caption{Population counts $P_{i}$}
	\end{subfigure}
	\begin{subfigure}[t]{0.3\textwidth}
		\centering
		\includegraphics[height=2.2in]{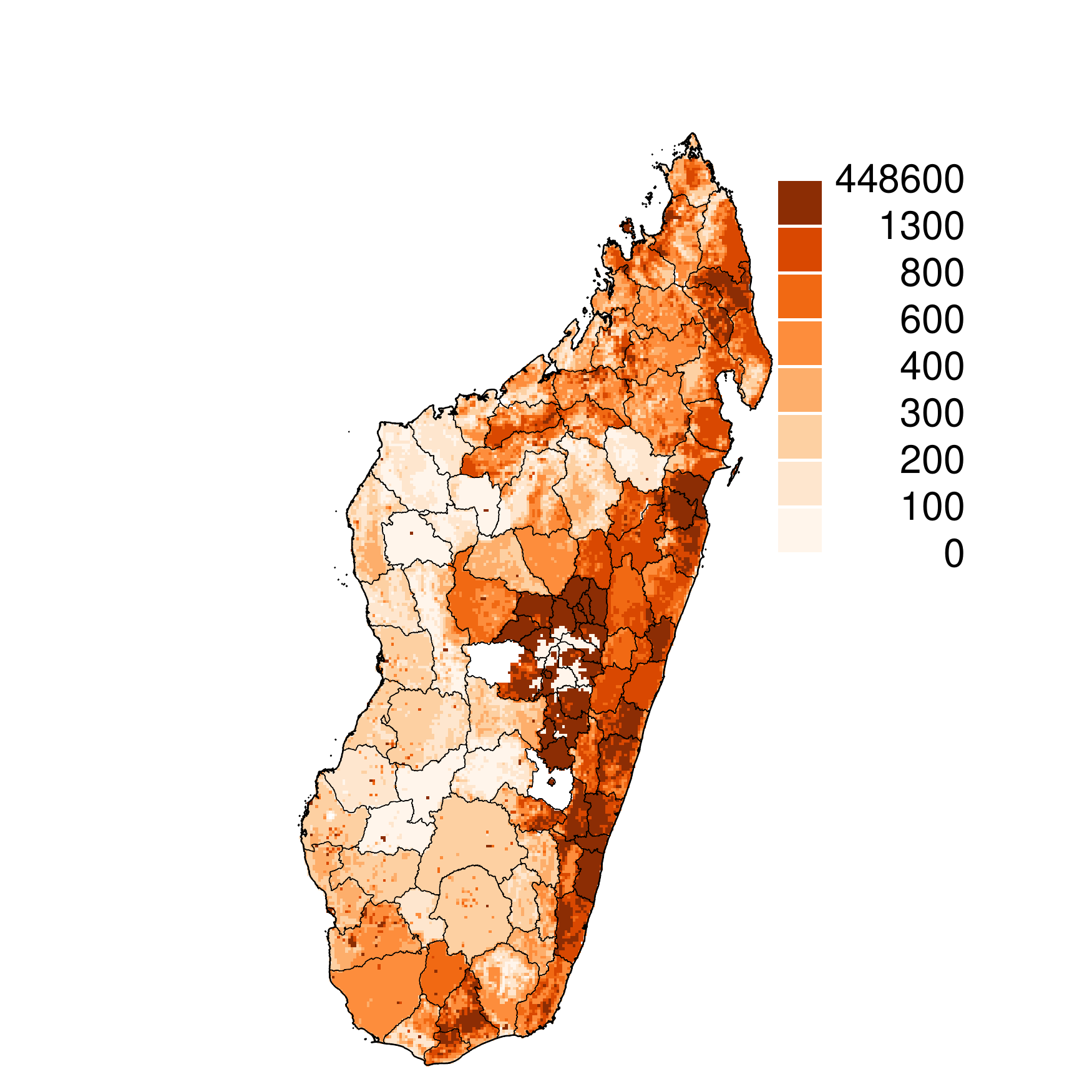}
		\caption{Population raster $P(\Q_{t})$}
		\label{fig:popfine}
	\end{subfigure}
	\begin{subfigure}[t]{0.3\textwidth}
		\centering
		\includegraphics[height=2in]{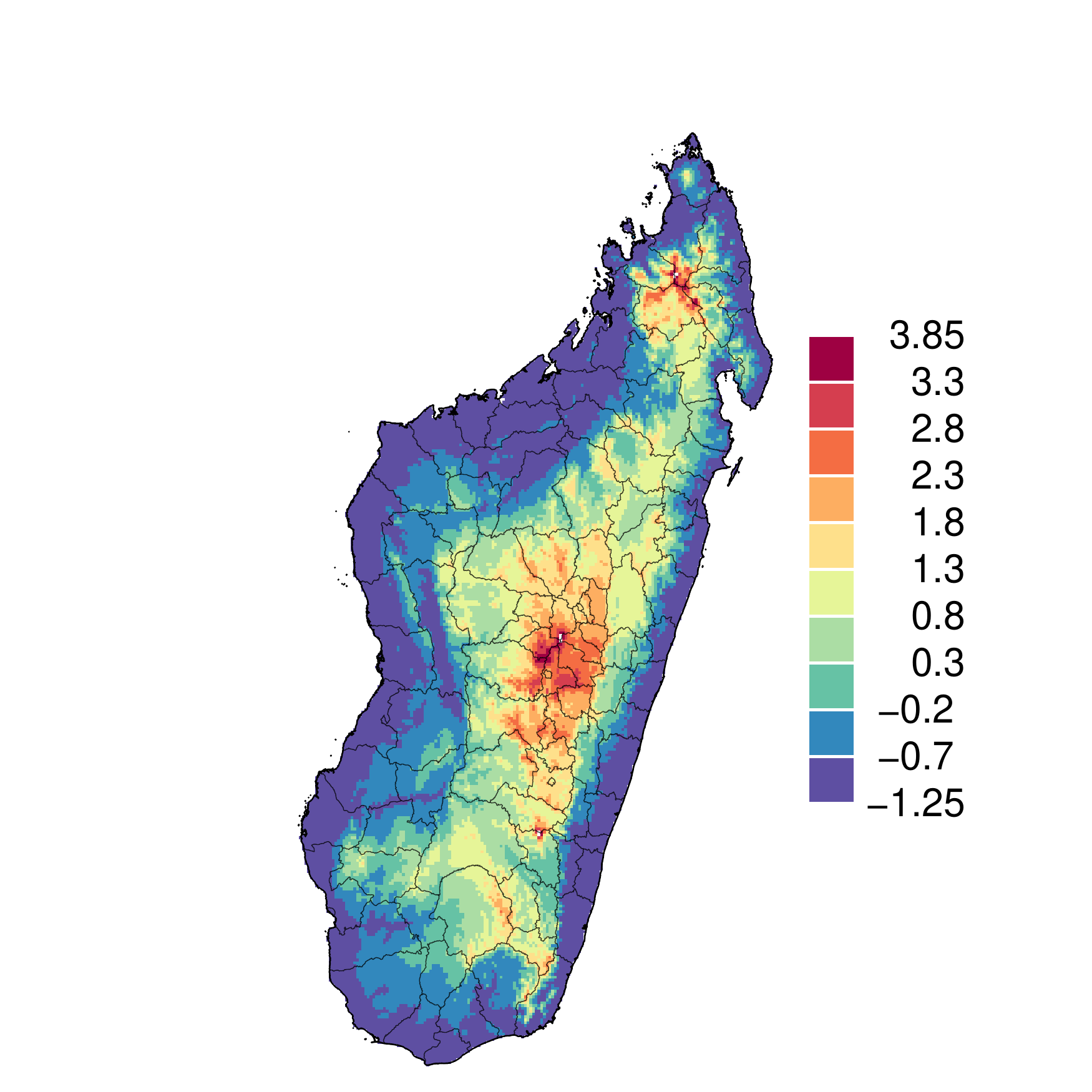}
		\caption{Elevation $x_{1}(\Q_{t})$}
	\end{subfigure}
	\begin{subfigure}[t]{0.3\textwidth}
		\centering
		\includegraphics[height=2in]{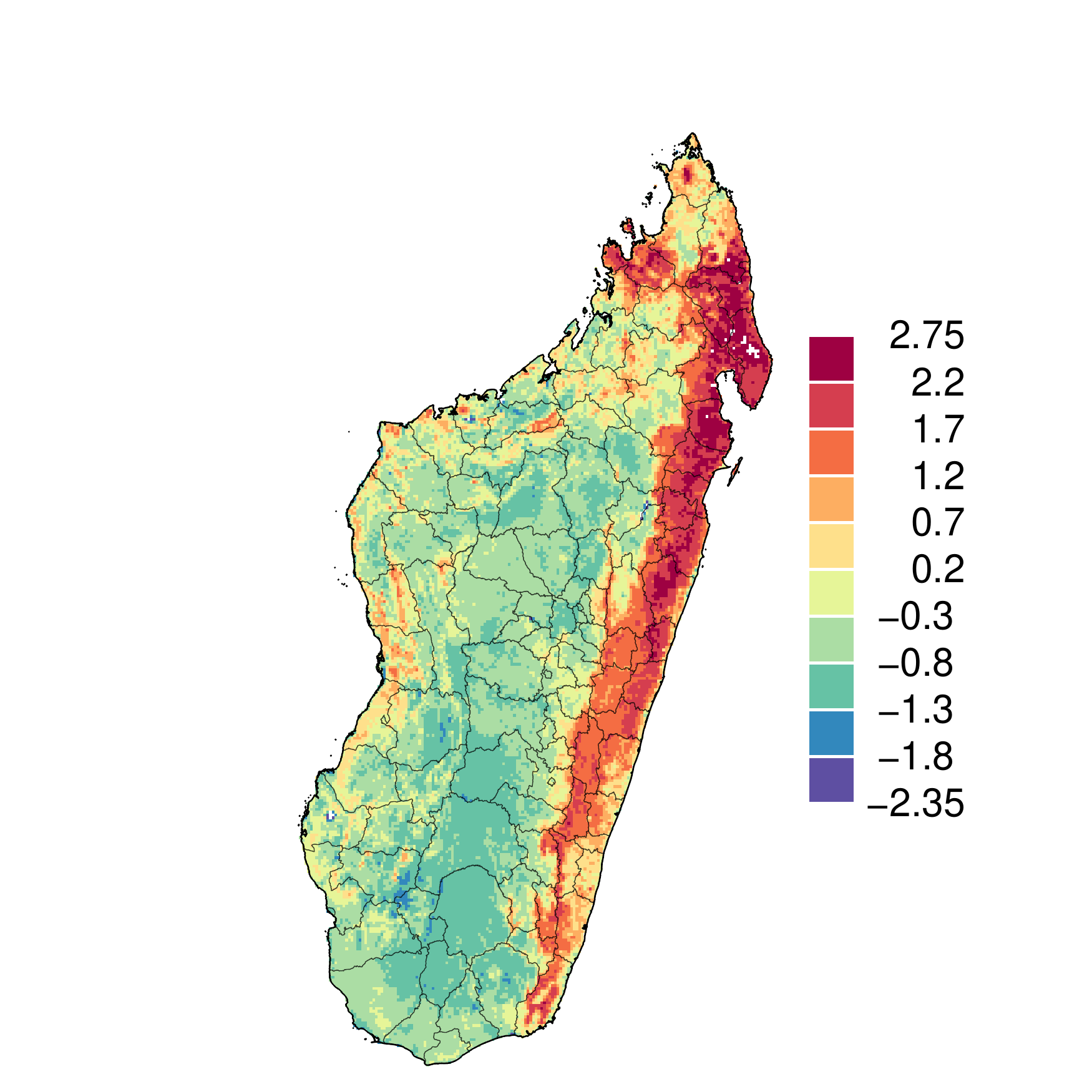}
		\caption{Vegetation index $x_{2}(\Q_{t})$}
	\end{subfigure}
	\begin{subfigure}[t]{0.3\textwidth}
		\centering
		\includegraphics[height=2in]{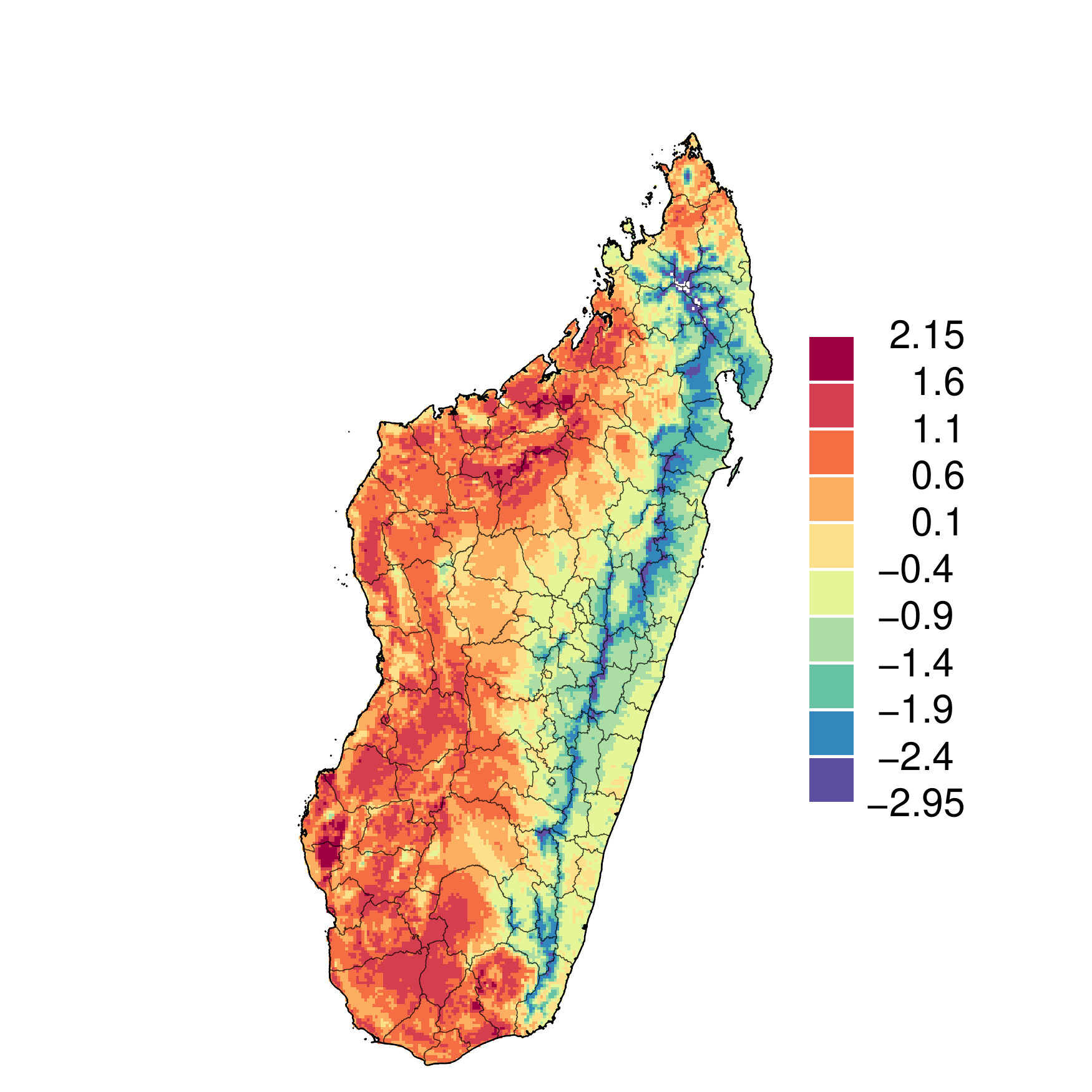}
		\caption{Surface temp. $x_{3}(\Q_{t})$}
	\end{subfigure}
	\caption{Case counts $Y_{i}$, population counts within each polygon $P_{i} = \int_{S_{i}}P(\mb{s})d\mb{s}$ and at the pixel level $P(\Q_{t})$, and covariate rasters $\mb{x}(\Q_{t}) = \left\{1,x_{1}(\Q_{t}),x_{2}(\Q_{t}),x_{3}(\Q_{t})\right\}, t\in[T]$ for Malaria incidence in Madagascar, \S\ref{subsec:aggspatial}.}
	\label{fig:mgcounts}
\end{figure}

The covariates are observed on a fine grid with cells $\Q_{t},t\in[T]$ such that $\cup_{t=1}^{T}\Q_{t}=\M$ and $\Q_{t}\cap\Q_{l}=\emptyset$ for $t\neq l\in[T]$. The population counts in each grid cell $P_{it} \equiv \int_{S_{i}\cap\Q_{t}}P(\mb{s})d\mb{s}$ are available (Figure \ref{fig:popfine}). Further approximate $x(\loc) \approx \mb{x}(\Q_{t})$ for $\loc\in\Q_{t}$, where $\mb{x}(\Q_{t})$ is shorthand for taking $\loc$ to be the centre of the cell $\Q_{t}$. Inferences for $\lambda(\cdot)$ are made on the grid $\Q_{t}$ by defining $\lambda(\loc) \approx \lambda(\Q_{t})$ for each $\mb{s}\in\Q_{t}$. Finally, define $\mb{\lambda} = \bracevec{\lambda(\Q_{t}):t\in[T]}$. Figure \ref{fig:mgcounts} shows the observed data at the polygon level and on the grid, as appropriate. The discrete model is then:
\*[
\datai_{i}|\mb{\lambda}_{\J_{i}},v_{i}&\ind\text{Poisson}\left( \exp(v_{i})\sum_{t\in\J_{i}}P_{it}\lambda(\Q_{t})\right), \ \J_{i} = \bracevec{t\in[T]: S_{i}\cap\Q_{t} \neq \emptyset}, \\
\eta_{t} &= \log\lambda(\Q_{t}) = \beta_{0} + \mb{x}(\Q_{t})\tpose\mb{\beta} + u(\Q_{t}), \ u(\Q_{t}) = \sum_{j=1}^{d}\basisfunction_{j}(\Q_{t})\uui_{j}, \\ 
\mb{U}|\mb{\theta}_{2} &\sim \text{Normal}\left[ \zero,\mb{\Sigma}(\paramsmallprior)\right], \ v_{i} \overset{iid}{\sim}\text{N}(0,\tau^{-1}), \ \tau^{-1/2}\sim \text{Exp}(\xi_{\tau}), \sigma\sim \text{Exp}(\xi_{\sigma}), \rho^{-1}\sim \text{Exp}(\xi_{\rho})
\]
where $\mb{\Sigma}^{-1}(\paramsmallprior)$ is the sparse approximation to the precision matrix of the weights $\uu$ on the mesh \citep{spde}. Following \citet{disaggregation} the nonlinear parameters $\paramsmall = (\log\tau,\log\sigma,\log\rho)$ are assigned priors such that $\tau^{-1/2}, \sigma$ and $\rho^{-1}$ follow independent Exponential distributions with parameters chosen to satisfy $\PP(\tau > 0.1) = \PP(\sigma > 0.1) = \PP(\rho < 3) = 0.01$ (which are Penalized Complexity priors of \citealt{pcprior} and \citealt{pcmatern}), as well as $\beta_{0} \sim \text{N}(0,2^2)$ and $\beta_{j}\sim\text{N}(0,0.4^2),j\in[p]$ independently.

\begin{figure}[p]
	\centering
	\begin{subfigure}[t]{0.45\textwidth}
		\centering
		\includegraphics[height=2in]{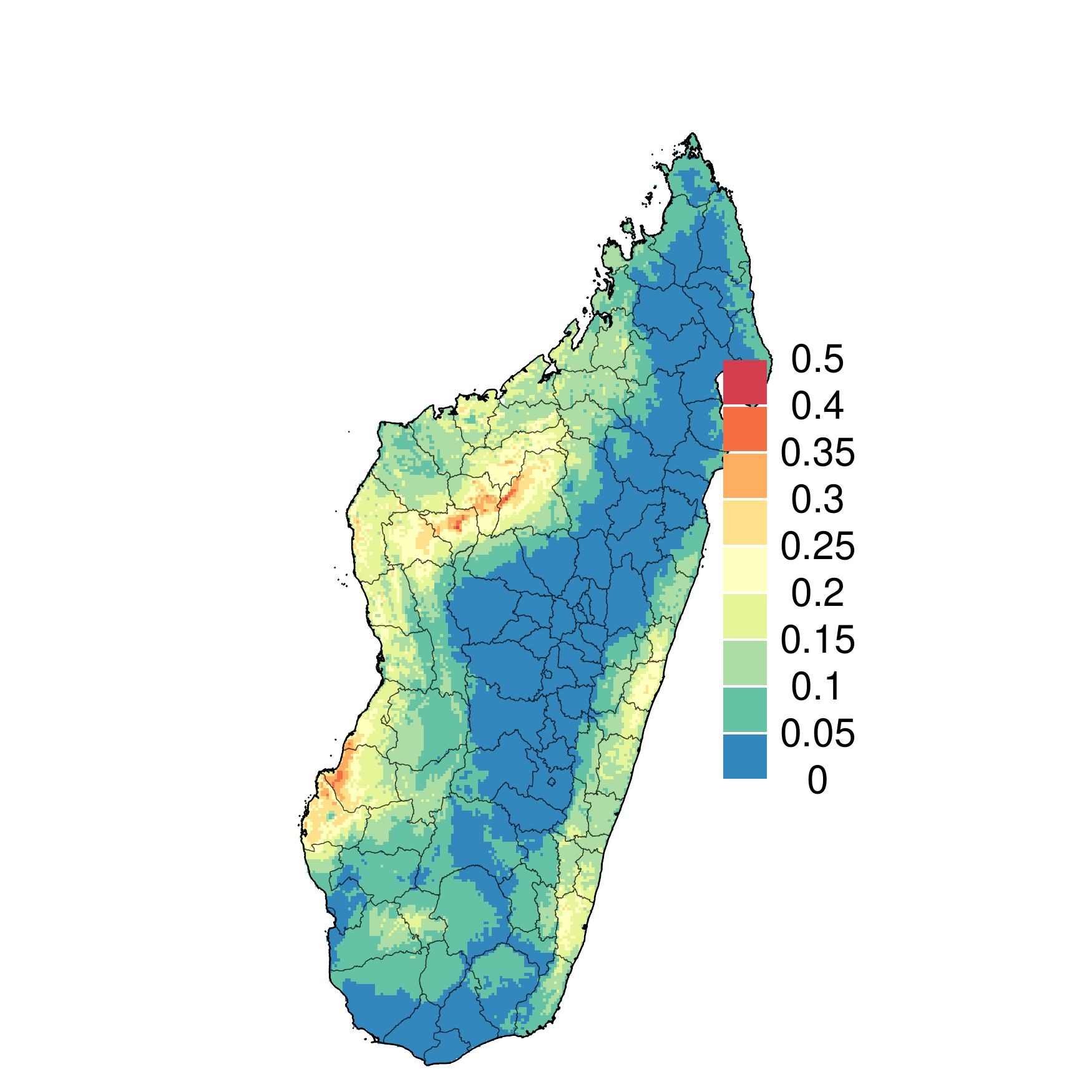}
		\caption{$\EE\left\{\lambda(\Q_{t}) | \mb{Y}\right\}$, ELGM}
	\end{subfigure}
	\begin{subfigure}[t]{0.45\textwidth}
		\centering
		\includegraphics[height=2in]{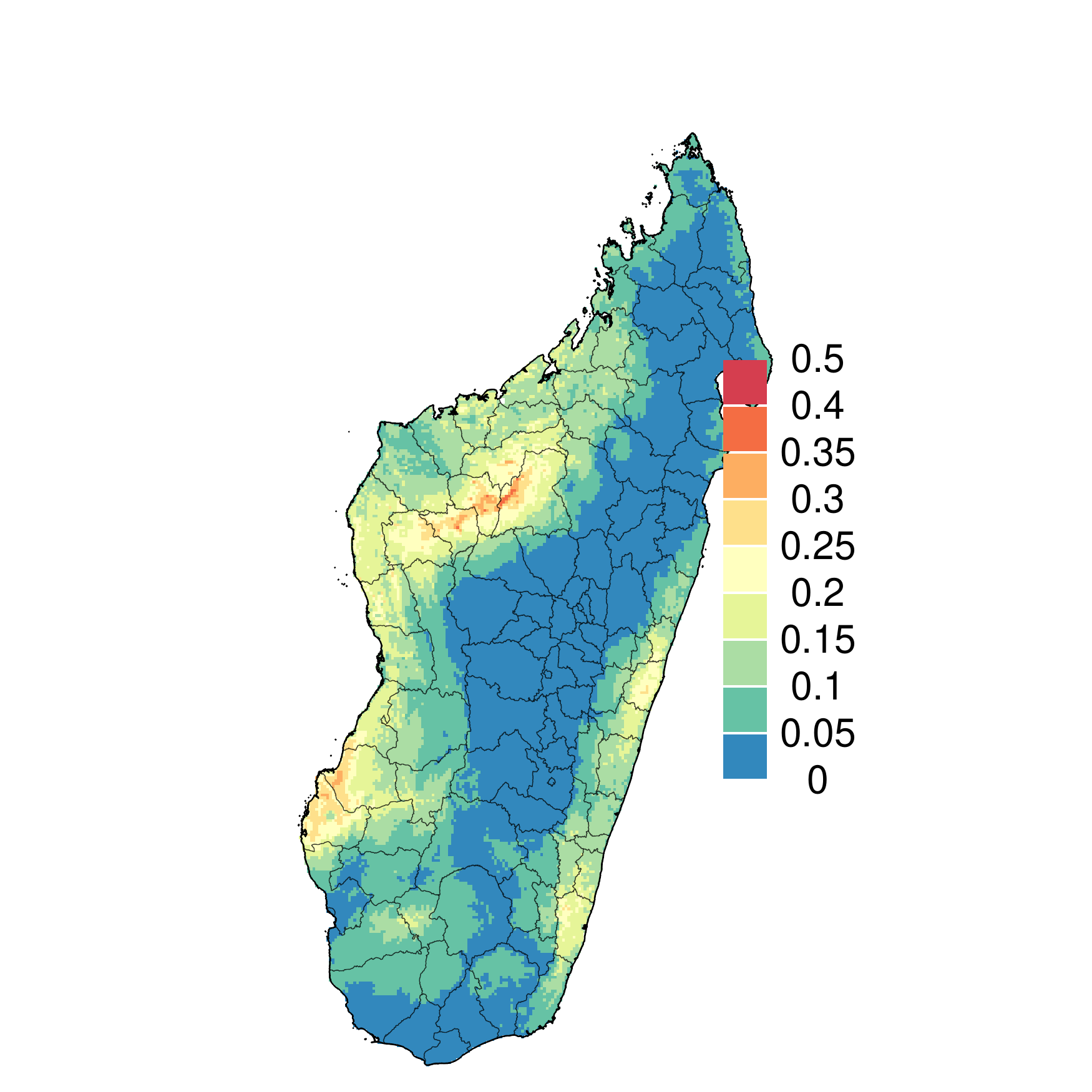}
		\caption{$\EE\left\{\lambda(\Q_{t}) | \mb{Y}\right\}$, MCMC}
	\end{subfigure}
	\begin{subfigure}[t]{0.45\textwidth}
		\centering
		\includegraphics[height=2in]{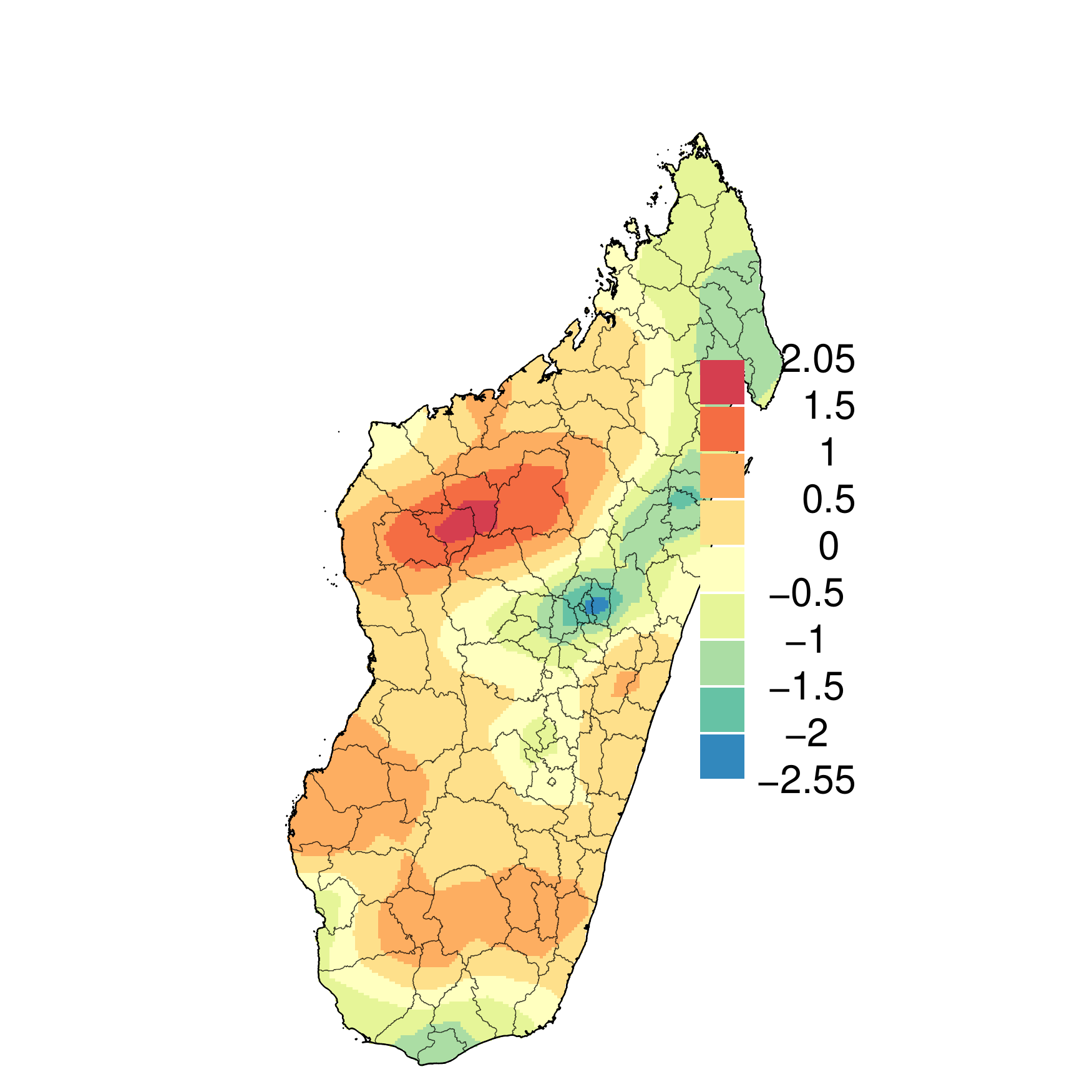}
		\caption{$\EE\left\{u(\Q_{t}) | \mb{Y}\right\}$, ELGM}
	\end{subfigure}
	\begin{subfigure}[t]{0.45\textwidth}
		\centering
		\includegraphics[height=2in]{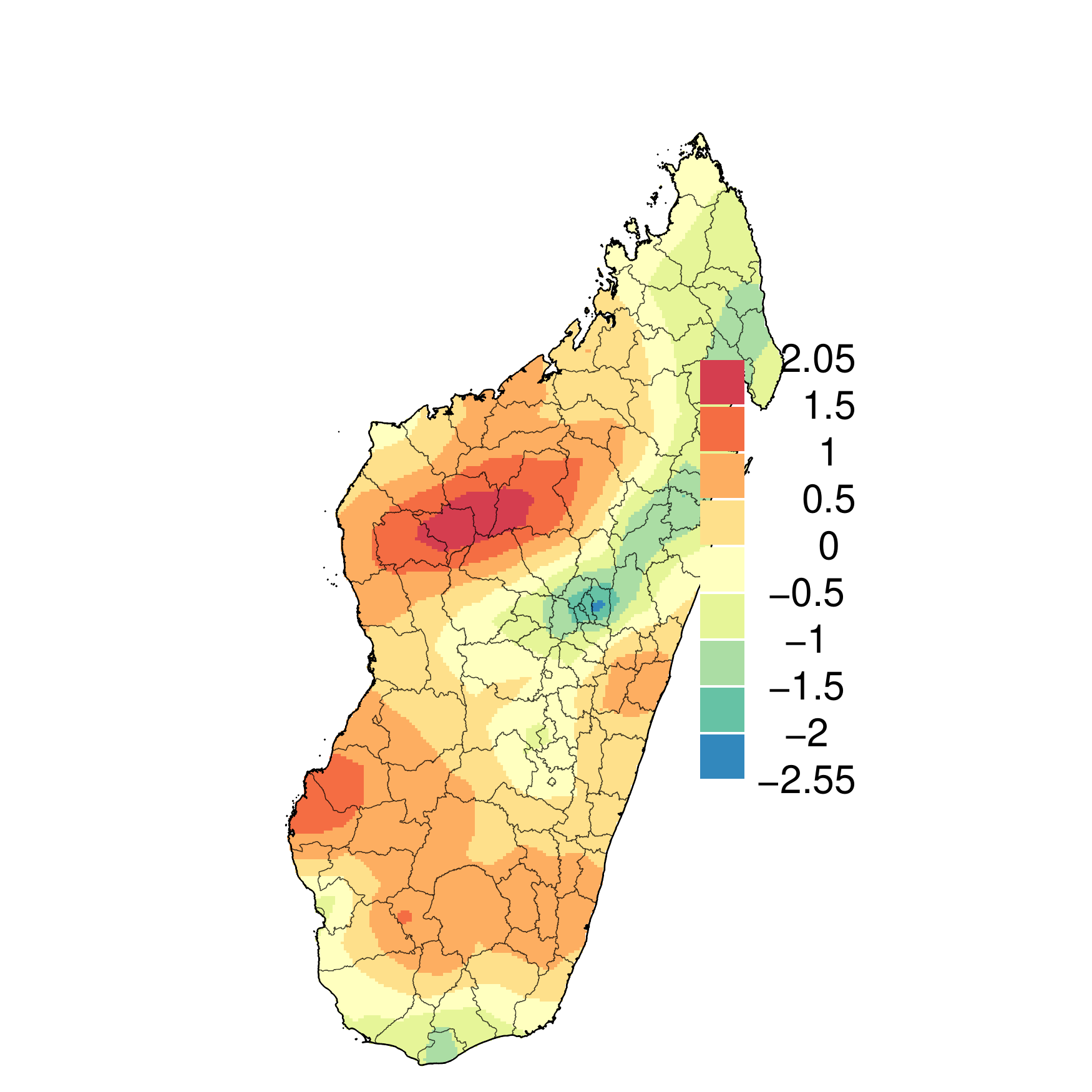}
		\caption{$\EE\left\{u(\Q_{t}) | \mb{Y}\right\}$, MCMC}
	\end{subfigure}
	\begin{subfigure}[t]{0.45\textwidth}
		\centering
		\includegraphics[height=2in]{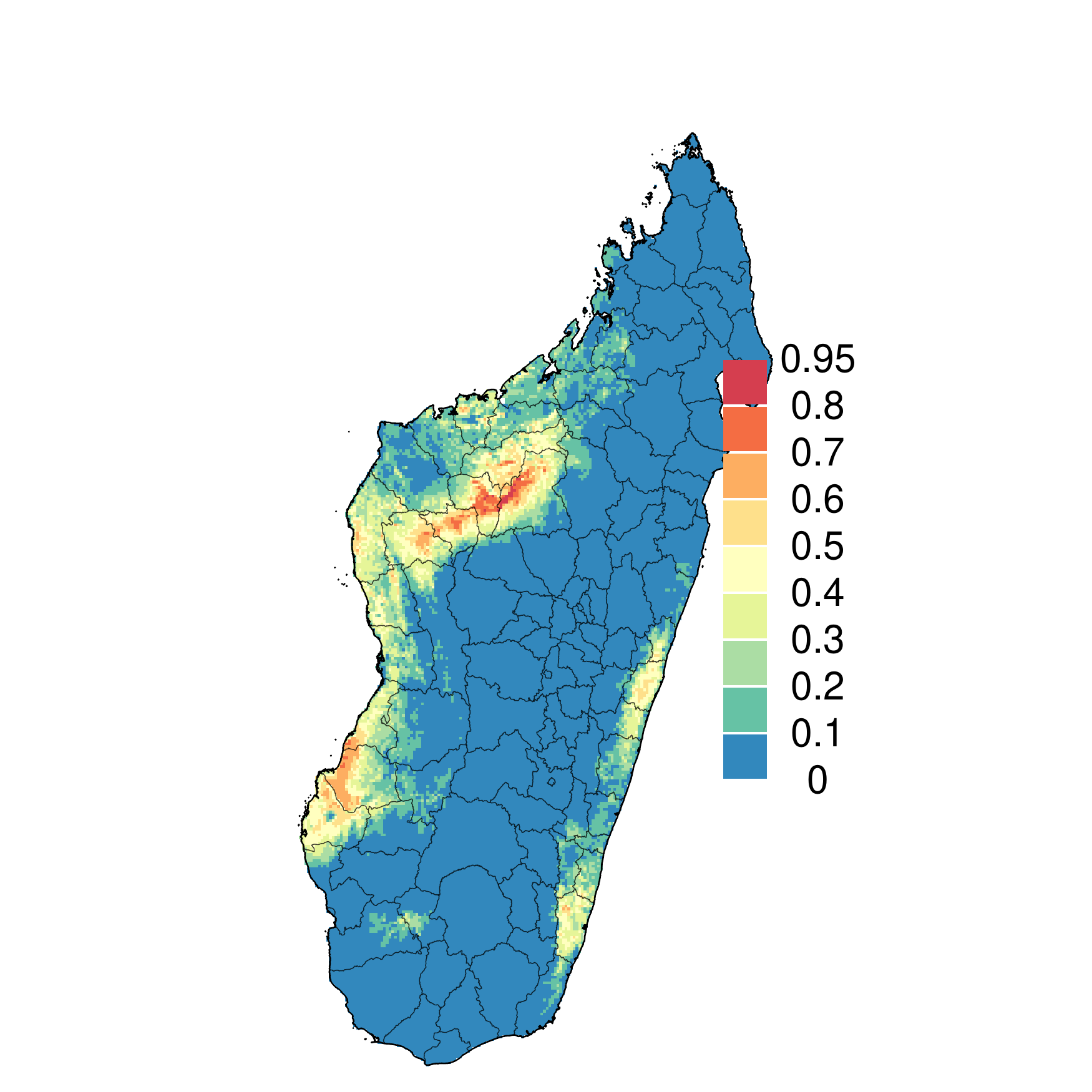}
		\caption{$\PP\left\{\lambda(\Q_{t}) > 0.2| \mb{Y}\right\}$, ELGM}
	\end{subfigure}
	\begin{subfigure}[t]{0.45\textwidth}
		\centering
		\includegraphics[height=2in]{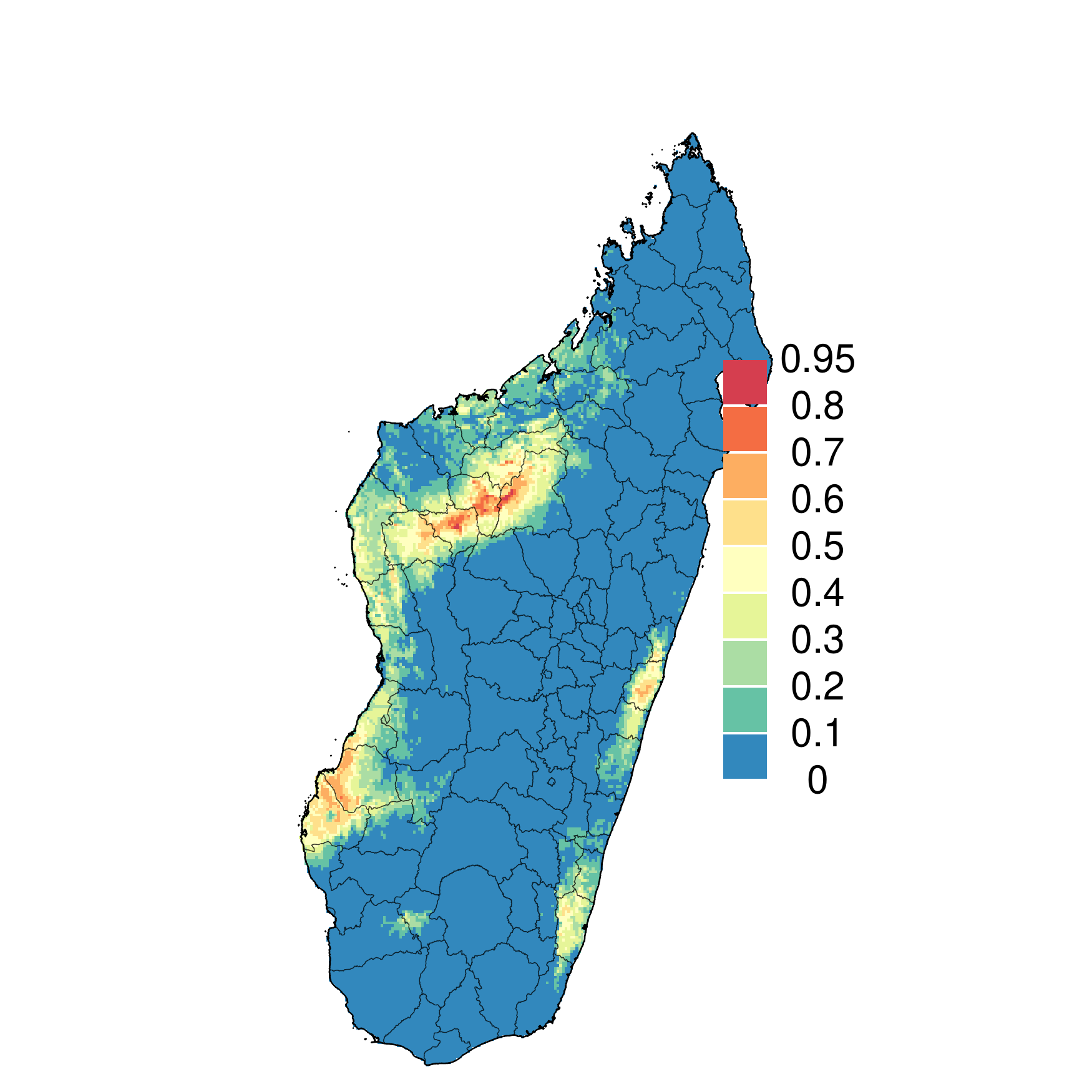}
		\caption{$\PP\left\{\lambda(\Q_{t}) > 0.2| \mb{Y}\right\}$, MCMC}
	\end{subfigure}
	\caption{Posterior summaries from the ELGM procedure and MCMC on the grid $\Q_{t},t\in[T]$, for the Malaria example of \S\ref{subsec:aggspatial}.}
	\label{fig:mgresults}
\end{figure}

\begin{figure}[p]
	\centering
	\begin{subfigure}[t]{0.32\textwidth}
		\centering
		\includegraphics[height=2in,width=2.3in]{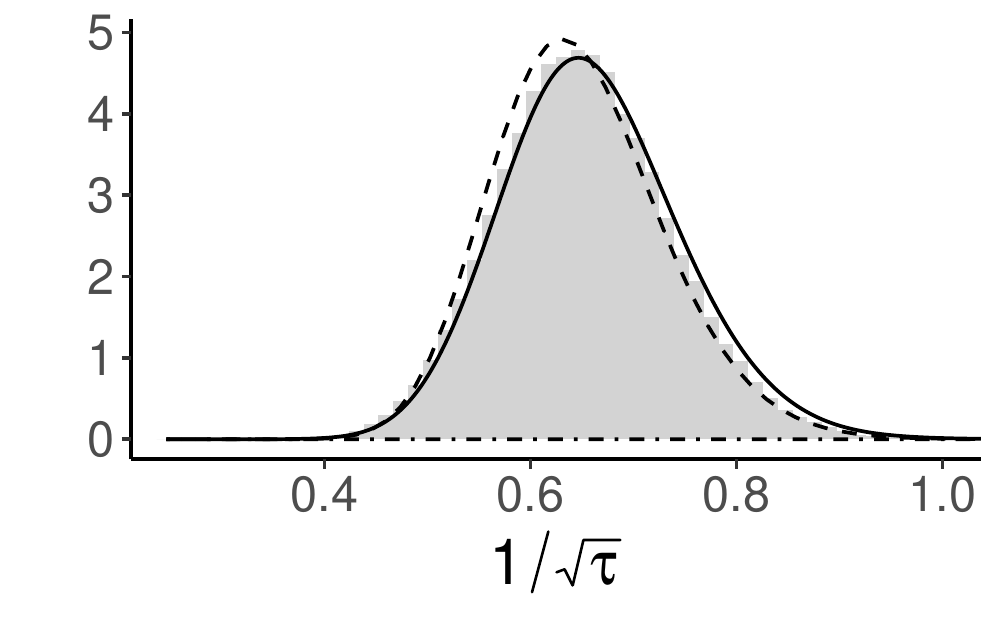}
		\caption{$\pi(\tau^{-1/2}|\mb{Y})$}
	\end{subfigure}
	\begin{subfigure}[t]{0.32\textwidth}
		\centering
		\includegraphics[height=2in,width=2.3in]{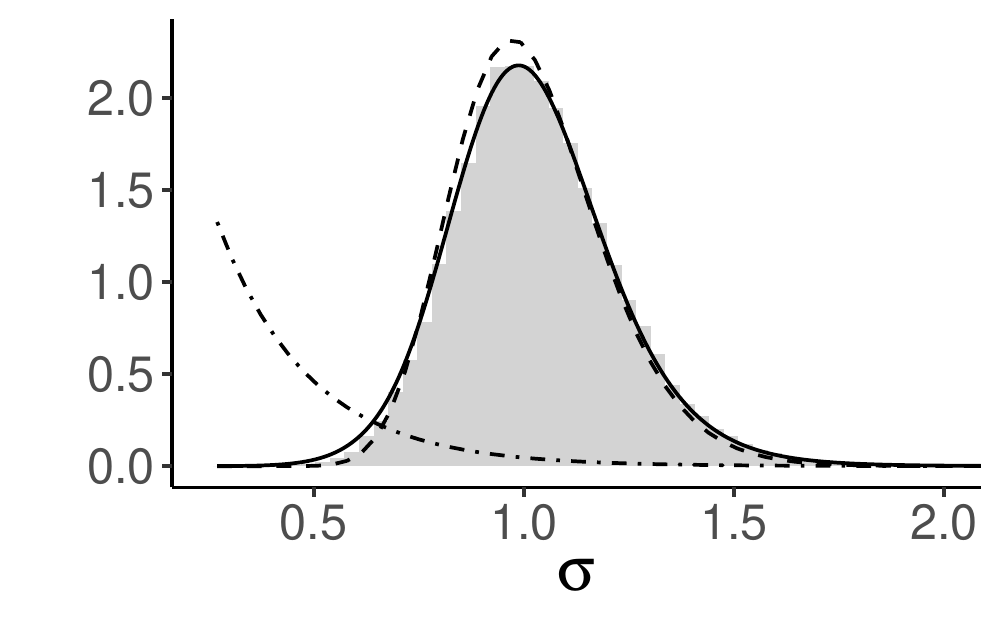}
		\caption{$\pi(\sigma|\mb{Y})$}
	\end{subfigure}
	\begin{subfigure}[t]{0.32\textwidth}
		\centering
		\includegraphics[height=2in,width=2.3in]{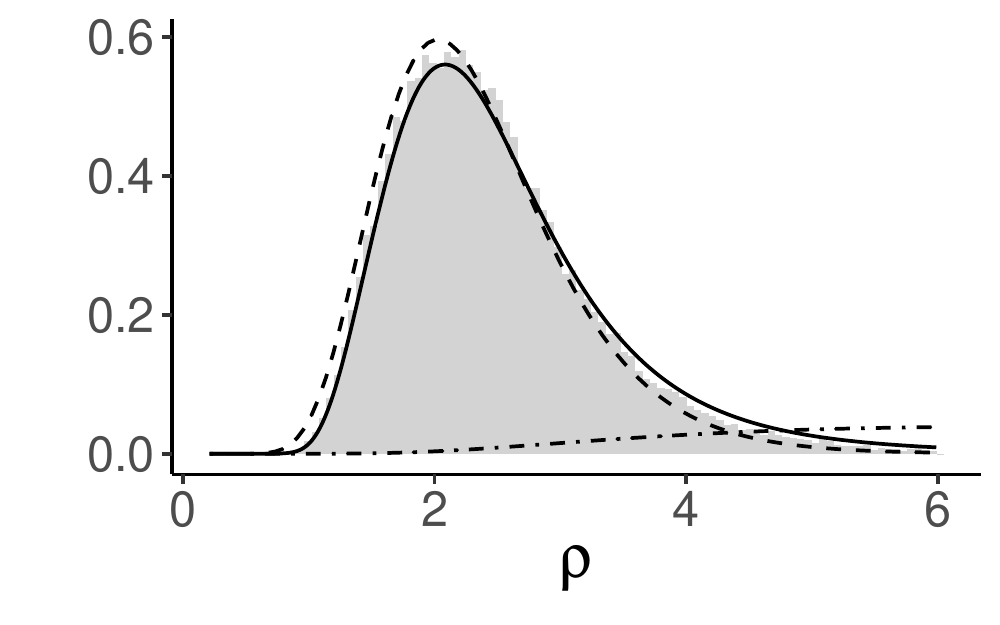}
		\caption{$\pi(\rho|\mb{Y})$}
	\end{subfigure}
	\caption{Approximate posteriors for $\tau^{-1/2}, \sigma, \rho$ using the ELGM procedure (---) and the Laplace approximation of \citet{disaggregation} (- - -), along with posterior samples from MCMC ($\textcolor{lightgray}{\blacksquare}$) and prior ($-\cdot-$), for the Malaria example of \S\ref{subsec:aggspatial}.}
	\label{fig:mgnonlinear}
\end{figure}

We fit this model using our ELGM approach (Algorithm \ref{alg:implementation}) with $k = 7$ quadrature points, adapting the excellent software of \citet{disaggregation} for use with the \texttt{aghq} package with a computation time of about $27$ minutes. We also fit the model using an MCMC run of $16$ chains with $5,000$ iterations (including a warmup of $1,000$) each for a total of $64,000$ usable iterations, using the \texttt{tmbstan} package \citep{tmbstan} at a wall computation time of approximately $29$ hours when running the chains in parallel. Figure \ref{fig:mgresults} shows the posterior mean incidence rates $\lambda(\cdot)$ obtained from $100$ independent posterior samples from the approximate posterior and from MCMC, as well as mean excess spatial variation $u(\cdot)$ and probabilty of exceeding $20\%$ risk, $\PP[\lambda(\cdot) > 0.2|\mb{Y}]$. Table \ref{tab:aggspatialcoef} shows the posterior mean and standard deviation for the regression coefficients and nonlinear parameters computed using the ELGM procedure and MCMC. Finally, Figure \ref{fig:mgnonlinear} shows the posteriors for $(\tau^{-1/2},\sigma,\rho)$ computed using our approach, the full Laplace approximation, and MCMC. The ELGM approach generally matches the MCMC output closely, and in the case of the posteriors for the nonlinear parameters (Figure \ref{fig:mgnonlinear}), appears more accurate than the full Laplace approximation of \citet{disaggregation}.

\begin{table}[p]
\centering
\begin{tabular}{|r|rr|rr|rr|rr|}
\hline
 & \multicolumn{2}{c|}{Mean} & \multicolumn{2}{c|}{SD} & \multicolumn{2}{c|}{$2.5\%$} & \multicolumn{2}{c|}{$97.5\%$} \\
Param. & ELGM & MCMC & ELGM & MCMC & ELGM & MCMC & ELGM & MCMC\\
\hline
$\beta_{0}$ & -3.054 & -3.15 & 0.363 & 0.348 & -3.75 & -3.84 & -2.29 & -2.48\\
$\beta_{\texttt{elev}}$ & -0.465 & -0.464 & 0.187 & 0.188 & -0.825 & -0.831 & -0.096 & -0.091\\
$\beta_{\texttt{vege}}$ & 0.385 & 0.384 & 0.204 & 0.206 & -0.017 & -0.016 & 0.784 & 0.791\\
$\beta_{\texttt{lst}}$ & 0.203 & 0.185 & 0.263 & 0.265 & -0.318 & -0.334 & 0.713 & 0.703\\
\hline
$\tau^{-1/2}$ & 0.660 & 0.652 & 0.088 & 0.0848 & 0.502 & 0.493 & 0.847 & 0.827\\
$\sigma$ & 1.02 & 1.03 & 0.198 & 0.194 & 0.668 & 0.693 & 1.45 & 1.46\\
$\rho$ & 2.57 & 2.47 & 0.984 & 0.833 & 1.32 & 1.31 & 5.00 & 4.54\\
\hline
\end{tabular}
\caption{Coefficient and nonlinear parameter posterior summaries from the ELGM procedure and MCMC, for the Malaria example of \S\ref{subsec:aggspatial}. Computation time for the ELGM was $27$ minutes, and for MCMC $29$ hours.}
\label{tab:aggspatialcoef}
\end{table}


\subsection{Spatial survival regression with partial likelihood}\label{subsec:coxph}

The Cox Proportional Hazards model with partial likelihood is a standard model for survival analysis problems. The partial (log) likelihood has a Hessian matrix which is completely dense, and hence is not compatible with the LGM framework, and is in fact the most computationally-intensive type of ELGM. We are unaware of any other approach that has been used to make Bayesian inferences based a Cox Proportional Hazards model with both a partial likelihood and a continuously-indexed latent spatial process, which we do in this example, and therefore this example serves to illustrate the breadth of our proposed method.

We consider a classic dataset of Leukaemia survival times in northern England where the goal is to infer the spatial variation in survival time. These data have been analyzed using survival models with parametric hazard functions \citep{leukaemia,spde}, resulting in simpler computations than when using partial likelihood. More recently \citet{inlasurvival} adapt the methodology of \citet{inla} to analyze these data using a semi-parametric hazard model which is still less computationally intensive than using partial likelihood, and where spatial variation is restricted to occur between predefined geographical regions. Unlike these previous analyses, we fit a Cox Proportional Hazards model using partial likelihood and a continuously-varying spatial model using the exact observed point locations, which is feasible in the ELGM framework introduced in this paper.

The data consist of the survival times of $n = 1043$ Leukaemia patients in northern England. Of these $n_{0} = 879$ were observed to die during the study period and $n - n_{0} = 164$ were right-censored. Denote the study region by $\M\subset\R^{2}$. Each subject has a point location of residence $\loc_{i}\in\M$, and covariate vector $\covx_{i}\in\R^{p}$ with $p = 4$ containing age, sex, white blood cell count (WBC), and the Townsend Deprivation Index (TPI), a measure of social deprivation. Let $\datai_{i},i\in[n]$ denote the survival time of the $i^{th}$ subject. For convenience suppose $0 < \datai_{1} < \cdots < \datai_{n_{0}}$ are observed and $(\datai_{i})_{i = n_{0}+1,\ldots,n}$ are censored. We use the following hierarchical model:
\*[
\pi\left(\data|\mb{\lambda}\right) &= \prod_{i=1}^{n_{0}}\frac{\lambda_{i}}{\sum_{j=i}^{n}\lambda_{j}}, \ \eta_{i} = \log\lambda_{i} = \beta_{0} + \covx_{i}^{T}\mb{\beta} + u(\loc_{i}), i\in[n], \ \J_{i} = \bracevec{i,\ldots,n}, \\
u(\loc) &\sim \GP\left\{0,\Matern_{\nu}(\cdot;\sigma,\rho)\right\}, \ \text{Cov}[u(\loc + \mb{h}),u(\loc)] = \Matern_{\nu}(\norm{\mb{h}};\sigma,\rho), \loc\in\M, \mb{h}\in\R^{2}.
\]
Because $\J_{1} = [n]$ we have $k,l\in\J_{1}$ for every $k,l\in[n]$ and hence $\llhesseta(\addpred,\paramsmalllik)$ is a fully dense matrix (\S\ref{sec:elgms}).

For the spatially-varying Gaussian process $u(\cdot)$ we use a Matern covariance function as in \S\ref{subsec:aggspatial}, however we do not use a piecewise constant approximation to $u(\cdot)$ with the sparse approximation to its precision matrix, and instead evaluate the dense covariance matrix at the observed point locations. The unknown spatial parameter is therefore $\uu = (\uui(\mb{s}_{i}))_{i\in[n]}$ with dimension $d = n$. The model for $\uu$ is:
\*[
\uu|\paramsmallprior &\sim \Normal\left\{\zero,\mb{\Sigma}(\paramsmallprior)\right\}, \ \mb{\Sigma}_{jk} = \Matern_{\nu}(\norm{\loc_{j} - \loc_{k}};\sigma,\rho), j,k\in[n],
\]
where, as in \S\ref{subsec:aggspatial}, $\Matern_{\nu}(\cdot;\sigma,\rho)$ is a Matern covariance function with fixed shape $\nu = 1$, and standard deviation $\sigma$ and range $\rho$ in the parametrization of \citet{diseasemapping}. The nonlinear parameters are $\paramsmall = (\log\sigma,\log\rho)$. Again as in \S\ref{subsec:aggspatial} we follow \citet{pcmatern} and choose independent Exponential priors on $\sigma$ and $\rho^{-1}$ which satisfy $\PP(\sigma > 1) = \PP(\rho < 20\text{km}) = 0.5$.

To approximate the joint predictive distribution $\uu^{*}|\data$ of $\uu^{*} = (u(\loc^{*}_{l}))_{l\in[L]}$ 
for any new locations $(\loc^{*}_{l})_{l\in[L]}\subset\M, L\in\N$, we make use of the full joint posterior approximation by first drawing $\parambig = (\uu,\mb{\beta})$ from $\approxpi(\parambig|\data)$ according to Algorithm \ref{alg:implementation}. We then draw from $\uu^{*}|\uu$ using methods implemented in the \texttt{RandomFields} package \citep{randomfields}.

\begin{figure}[p]
\centering
	\begin{subfigure}[t]{0.49\textwidth}
	\centering
	\includegraphics[height=3in]{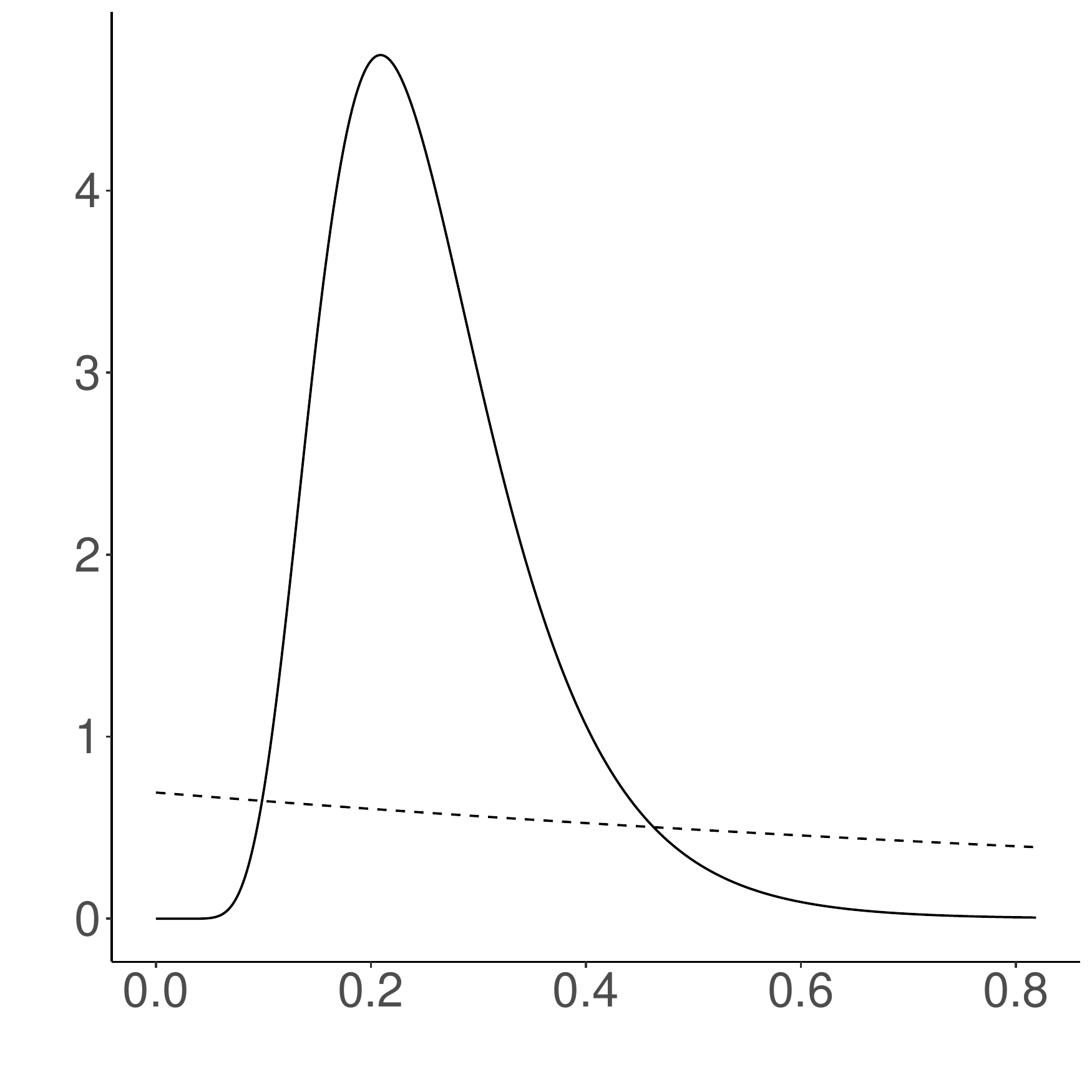}
	\caption{$\widetilde{\pi}_{\LA}(\sigma|\data)$}
	\end{subfigure}
	\begin{subfigure}[t]{0.49\textwidth}
	\centering
	\includegraphics[height=3in]{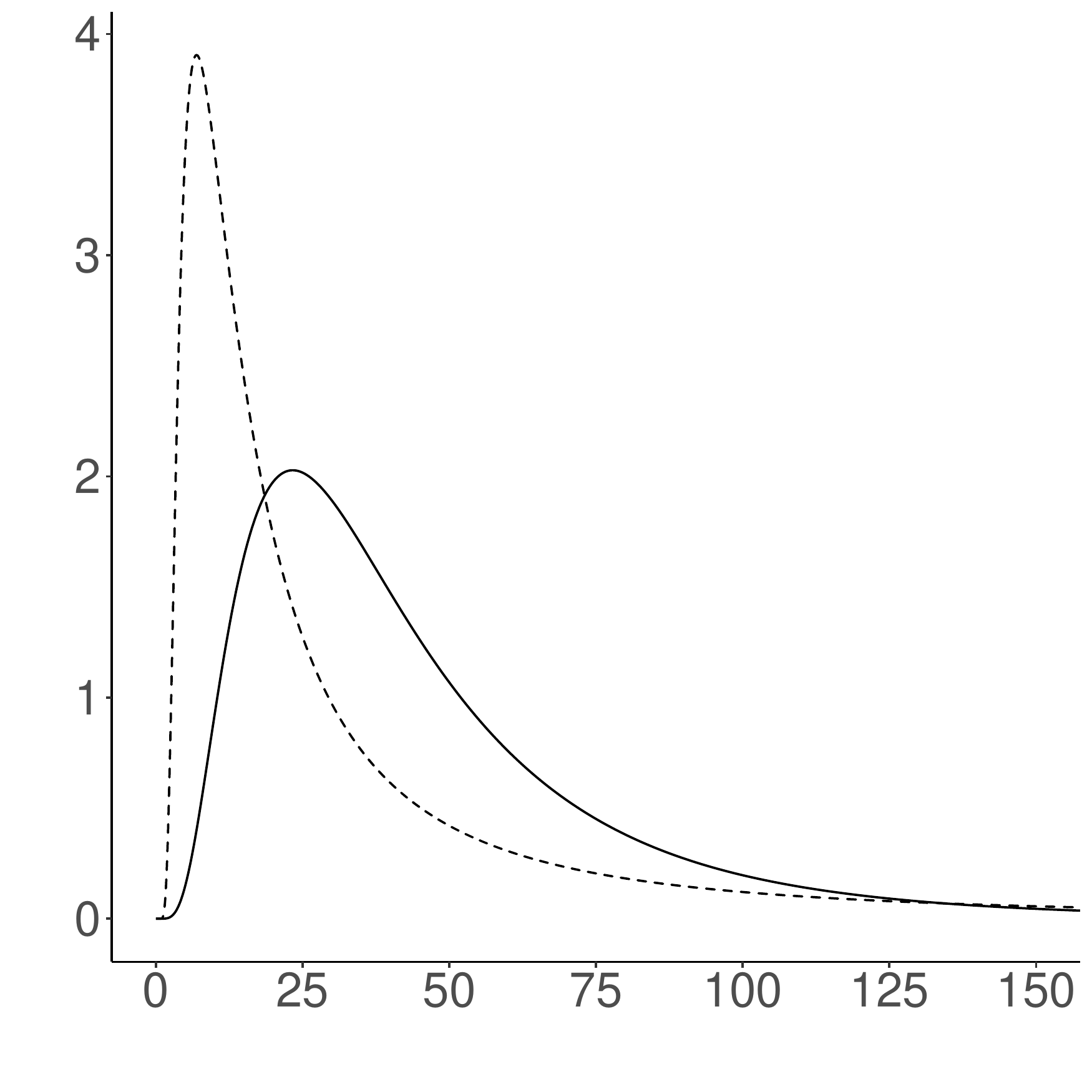}
	\caption{$\widetilde{\pi}_{\LA}(\rho|\data)$}
	\end{subfigure}
	\begin{subfigure}[t]{0.49\textwidth}
	\centering
	\includegraphics[height=3in]{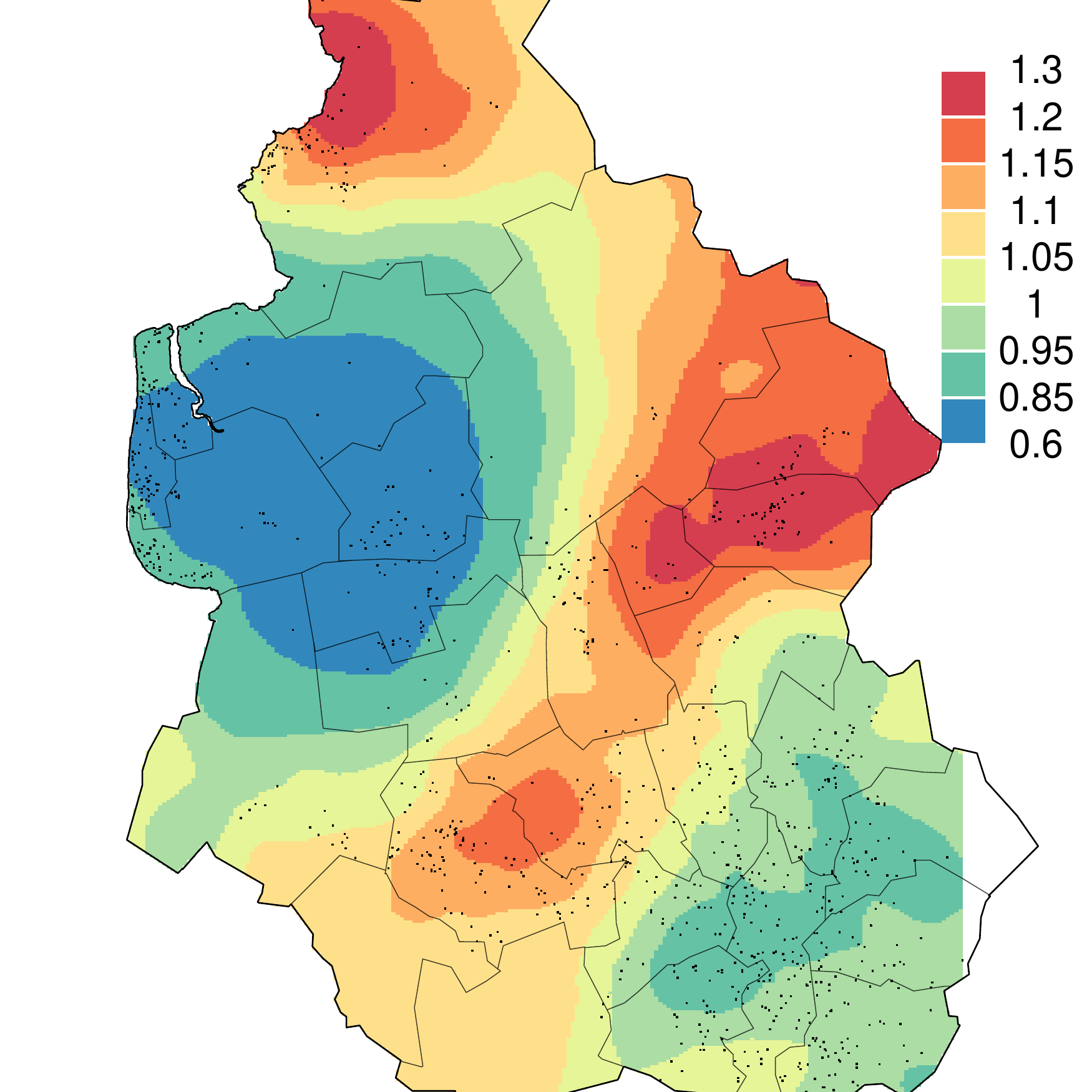}
	\caption{$\EE[\exp \bracevec{u(\loc)}|\data]$}
	\end{subfigure}
	\begin{subfigure}[t]{0.49\textwidth}
	\centering
	\includegraphics[height=3in]{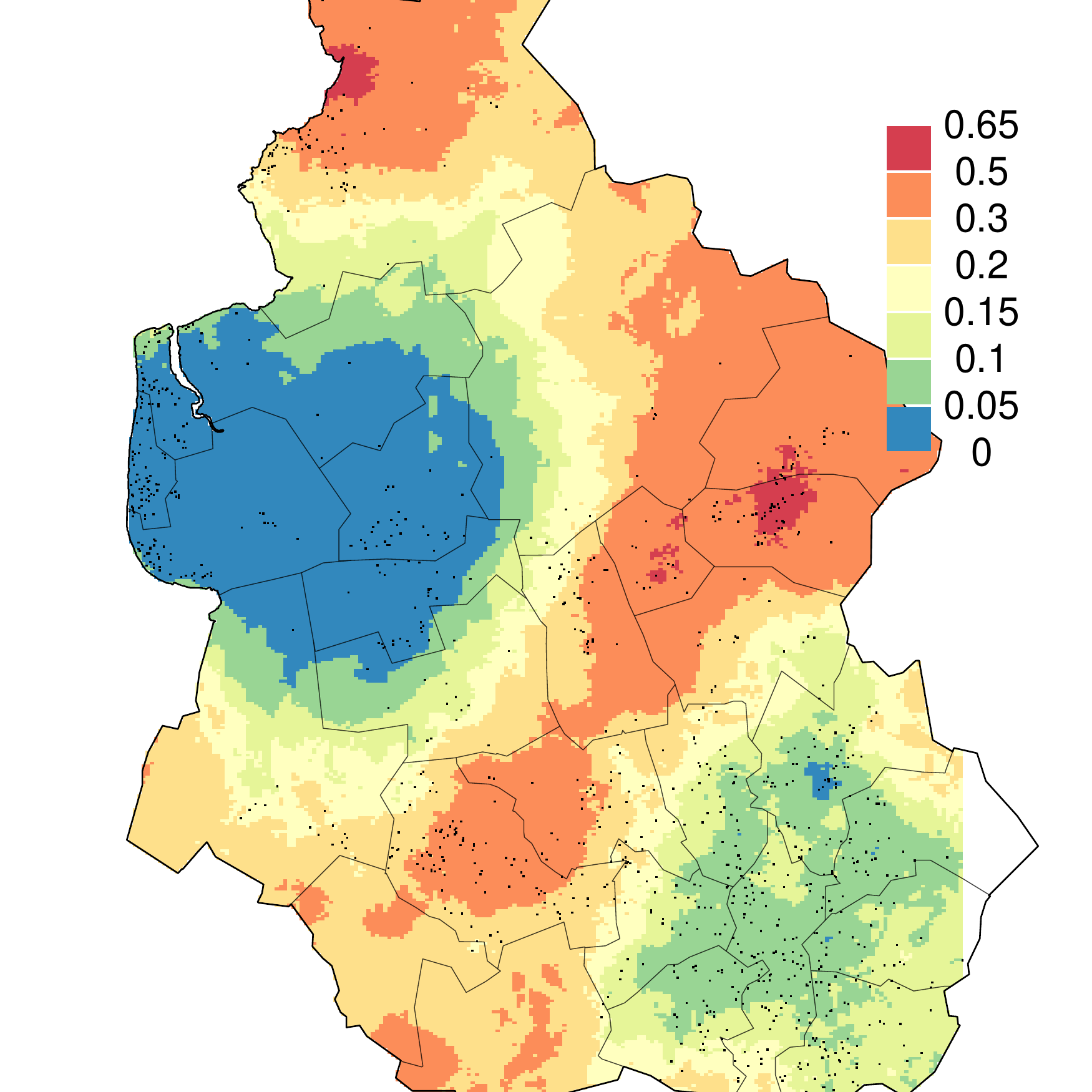}
	\caption{$\PP[\exp \bracevec{u(\loc)} > 1.2|\data]$}
	\end{subfigure}
\caption{(Top) Priors (- - -) and approximate posterior distributions (---) for (a) the spatial standard deviation and (b) range (KM, density $\times10^{5}$). (Bottom) maps showing (c) point locations of deaths $(\cdot)$ and approximate posterior mean excess spatial variation and (d) approximate posterior probability of exceeding $20\%$ excess spatial variation for the spatial survival analysis of \S\ref{subsec:coxph}}
\label{fig:coxph-results}
\end{figure}

Table \ref{tab:coxph} shows the posterior medians and credible regions for the coefficients of age, sex, Cell Count (WBC) and Deprivation (TPI), along with those for the the nonlinear parameters. The use of partial likelihood means that these are obtained without assumptions on the form of the baseline hazard. Figure \ref{fig:coxph-results} shows the approximate posterior distributions of the two nonlinear parameters in the Matern covariance function as well as the approximate posterior mean and $20\%$ exceedence probabilities obtained from $100$ draws from $\uu^{*}|\data$ for a fine grid of points $\loc^{*}_{1},\ldots,\loc^{*}_{L}, L = 58,215$ obtained by laying a $400\times 200$ square over the study area and taking the intersection of this square with the map. The use of exact point locations and simulation on a fine grid leads to a higher-resolution estimate than that of \citet{leukaemia} and \citet{spde}, though the overall pattern appears to match that reported by these previous analyses.

\begin{table}
\caption{\label{tab:coxph} Posterior moments and quantiles for fixed effects and nonlinear parameters for the Leukaemia data of Example \ref{subsec:coxph}.}
\centering
\begin{tabular}{|l|rrrr|}
\hline
Variable & Mean & SD & $2.5\%$ & $97.5\%$ \\
\hline
Age & 0.029 & 0.002 & 0.025 & 0.033\\
Male & 0.029 & 0.070 & -0.107 & 0.171\\
Cell Count & 0.005 & 0.001 & 0.003 & 0.004\\
Deprivation & 0.027 & 0.010 & 0.009 & 0.046\\
\hline
SD, $\sigma$ & 0.257 & 0.095 & 0.115 & 0.499\\
Range, $\rho$ (KM) & 44.2 & 32.0 & 9.77 & 134\\
\hline
\end{tabular}
\end{table}


\subsection{Estimating the mass profile of the Milky Way with measurement error}\label{subsec:astro}

In a series of papers, \citet{gwen2}, \citet{gwen3}, \citet{gwen5} and \citet{gwen4} develop and test a model for estimating the mass of the Milky Way Galaxy using position and velocity measurements of star clusters in its orbit. They use a physical model which implies a probability distribution for the position and velocity of star clusters, which depends on parameters that determine the mass of the Galaxy at any radial distance from its centre. This distribution is used in a hierarchical model which incorporates measurement uncertainties for the multivariate response, and missing data. Inference is based on posterior distributions for the parameters of interest, upon which estimates and uncertainty quantification for the mass of the Galaxy are based. 

Throughout these papers, posteriors are computed using MCMC which requires extensive user tuning \citep{gwen3}. Here we present similar results to \citet{gwen3} using the ELGM procedure with a running time faster than that reported by \citet{gwen3} and no user tuning, although we simplify some aspects of inference for purposes of illustration.

We describe only the relevant statistical details of this example; the physical details are well beyond the scope of this paper. Let $\data_{i} = (\datai_{i,1},\datai_{i,2},\datai_{i,3},\datai_{i,4})$ denote the four kinematic measurements taken for the $i^{th}$ star cluster: position, line-of-sight velocity, proper motion in right-ascention corrected for declination, and proper motion in declination, and let $\data = (\data_{i})_{i\in[n]}$. These {\it heliocentric} measurements are taken with respect to the position and motion of the sun, and are converted to {\it galactocentric} position and motion relative to the centre of the Galaxy using a deterministic, nonlinear transformation $\Omega(\cdot)$. The probability distribution relating the measurements to the mass of the Galaxy is defined in this galactocentric frame of reference. Prior to transformation, each heliocentric measurement is subject to a random measurement error $\uu_{i}$ with mean zero and fixed standard deviations $\mb{\delta}_{i}$ which are reported as part of the measurement process. 

For simplicity we consider only clusters with complete kinematic measurements and include measurement errors on the position and line-of-sight velocity only, defining $\uu_{i} = (\uui_{i1},\uui_{i2}),\parambig = (\uu_{i})_{i\in[n]}\in\R^{2n}$, and letting $\data_{i}^{*} = (\datai_{i,1} + \uui_{i,1},\datai_{i,2} + \uui_{i,2},\datai_{i,3},\datai_{i,4})$ for $i\in[n]$. There are $n = 70$ star clusters with complete data. A probability distribution $\pi\left\{\Omega(\mb{y}_{i})|\paramsmalllik\right\}$ over the galactocentric measurements $\Omega(\mb{y}_{i}),i\in[n]$ depends on parameters $\paramsmalllik = (\Psi_{0},\gamma,\alpha,\beta)$, and the mass of the Galaxy at radial distance $r$ kiloparsecs (kpc) from its centre, $M_{\paramsmalllik}(r) = \gamma\Psi_{0} r^{1-\gamma}$, is the object of inferential interest. Parameter transformations and strongly-informative priors on $\paramsmalllik$ are set according to the extensive discussion in \citet{gwen2}. For further detail, refer to \citet{gwen2} and \citet{gwen3}.

The model is as follows:
\begin{equation}\begin{aligned}
\pi(\data | \parambig,\paramsmalllik) &= \prod_{i=1}^{n}\pi\left\{\Omega(\data_{i}^{*})|\paramsmalllik\right\}, \ \J_{i} = \bracevec{\bracevec{i,1},\bracevec{i,2}}, \ \uu_{i} \overset{ind}{\sim} \Normal\left\{\zero,\text{diag}(\mb{\delta}_{i})\right\}, i\in[n],
\end{aligned}\end{equation}
where $i \in[n]$. The measurement standard deviations $\mb{\delta}_{i} = (\delta_{i1},\delta_{i2}),i\in[n]$ are reported for each cluster and taken as fixed and known. The strongly-informative priors are $\Psi_{0}\sim\text{Unif}(1,200), \gamma\sim\text{Unif}(0.3,0.7),\alpha - 3 \sim\text{Gamma}(1,4.6)$ and $\beta\sim\text{Unif}(-0.5,1)$ \citep{gwen2}. The index sets $\J_{i},i\in[n]$ have $\abs{\J_{i}} = 2 > 1$ for each $i\in[n]$, because the observations are multivariate. This is therefore an example of an ELGM compatible with our methodology, but not an LGM.

\begin{figure}[t]
\centering
	\begin{subfigure}[t]{0.49\textwidth}
	\centering
	\includegraphics[height=3in]{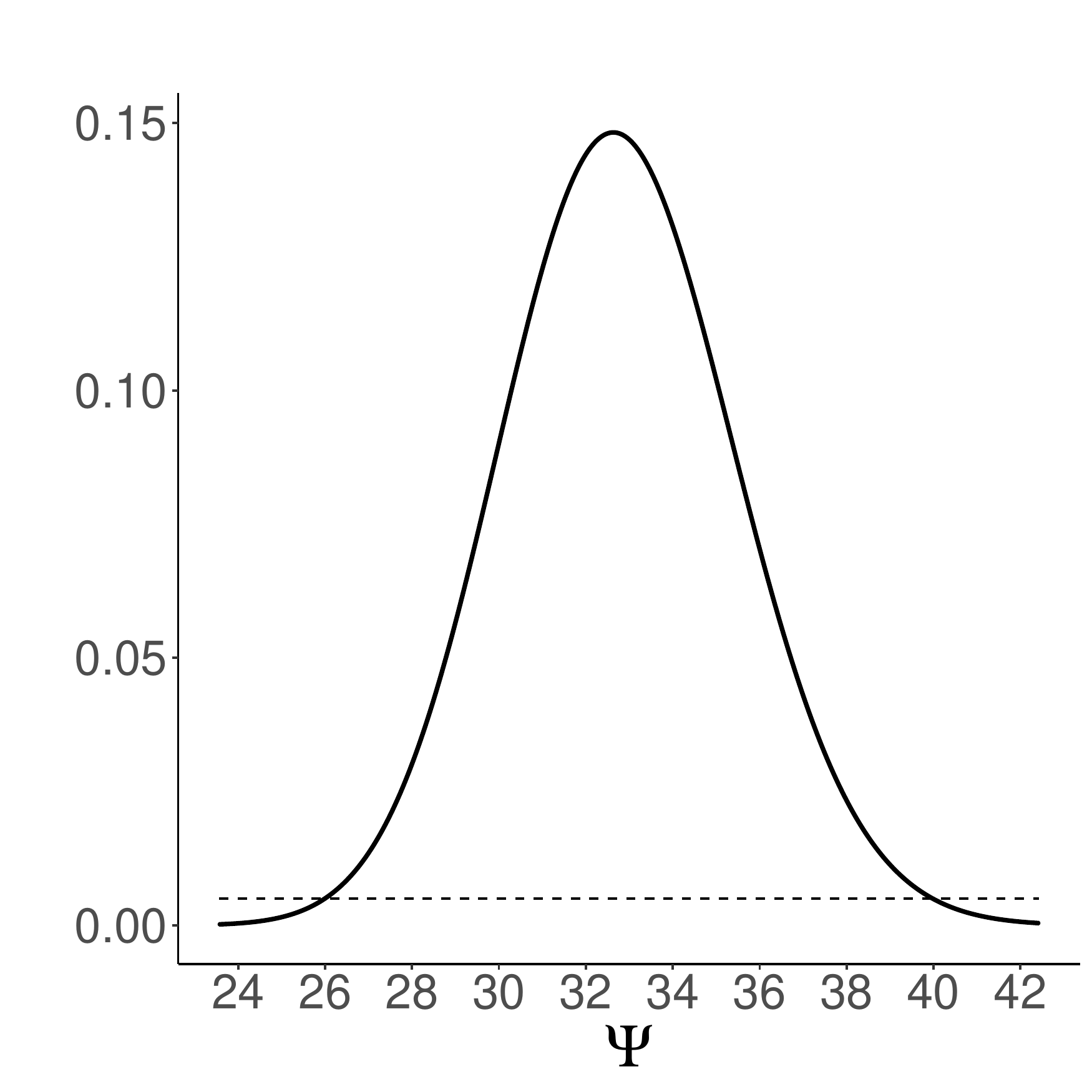}
	\caption{$\widetilde{\pi}_{\LA}(\Psi_{0}|\mb{Y})$}
	\end{subfigure}
	\begin{subfigure}[t]{0.49\textwidth}
	\centering
	\includegraphics[height=3in]{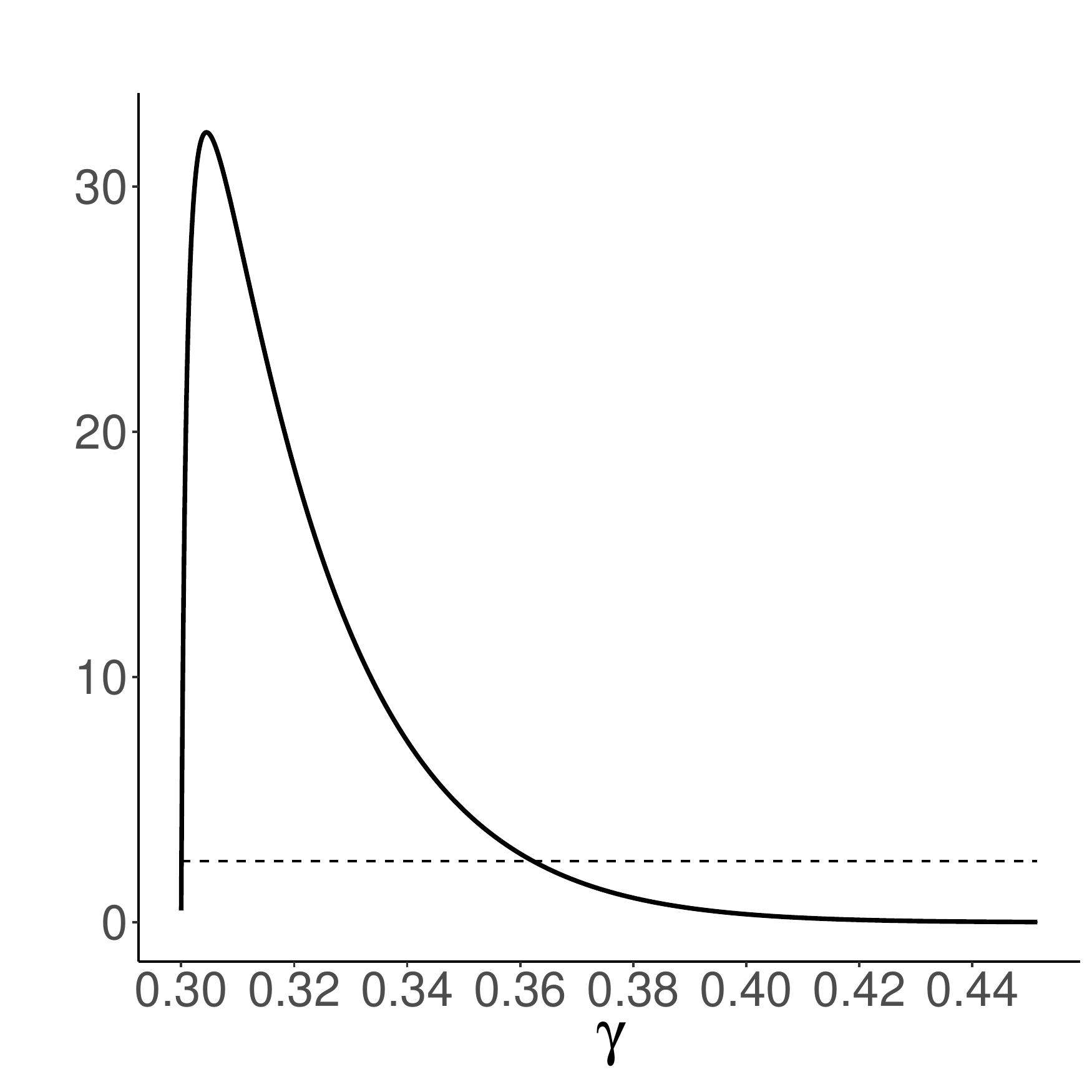}
	\caption{$\widetilde{\pi}_{\LA}(\gamma|\mb{Y})$}
	\end{subfigure}
	\begin{subfigure}[t]{0.49\textwidth}
	\centering
	\includegraphics[height=3in]{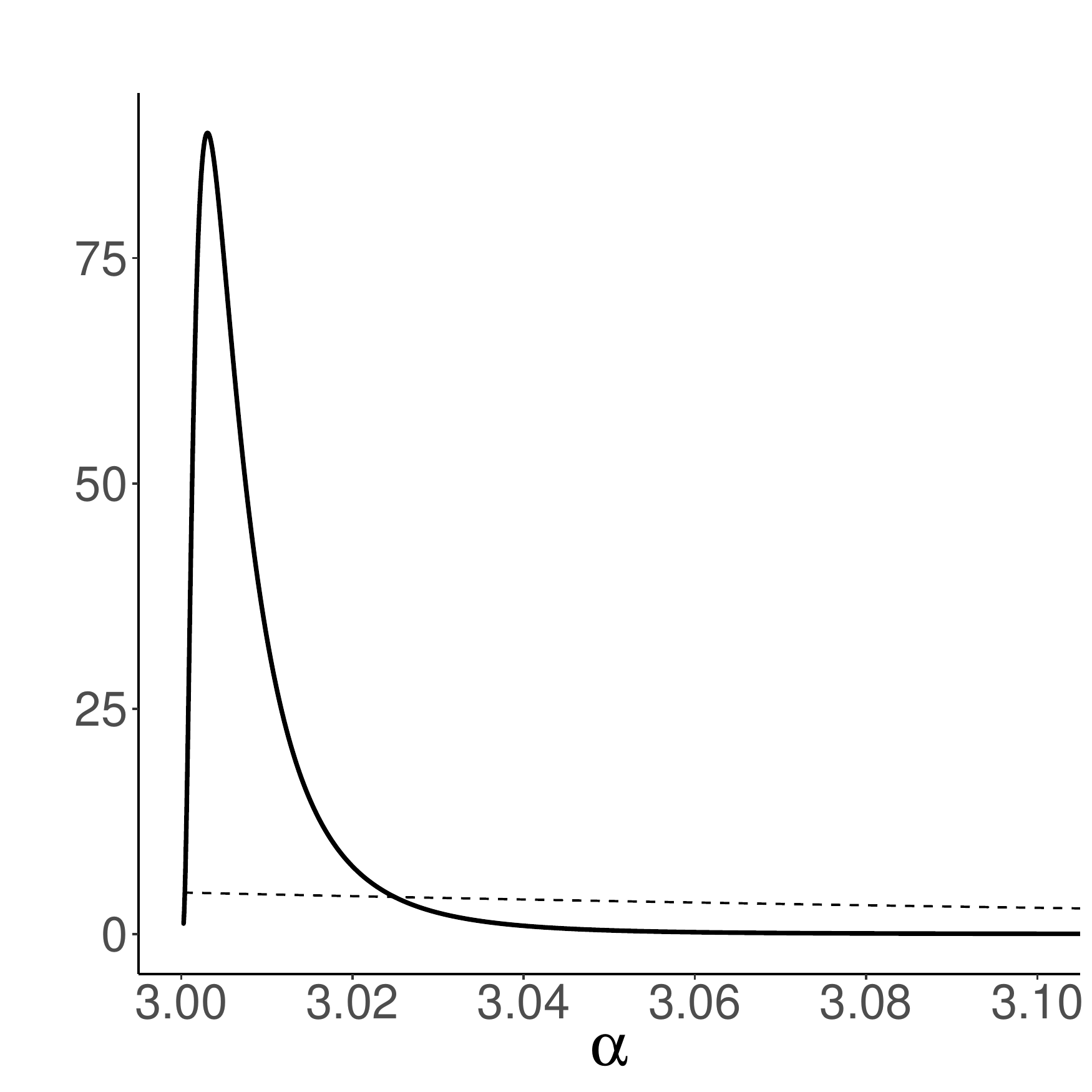}
	\caption{$\widetilde{\pi}_{\LA}(\alpha|\mb{Y})$}
	\end{subfigure}
	\begin{subfigure}[t]{0.49\textwidth}
	\centering
	\includegraphics[height=3in]{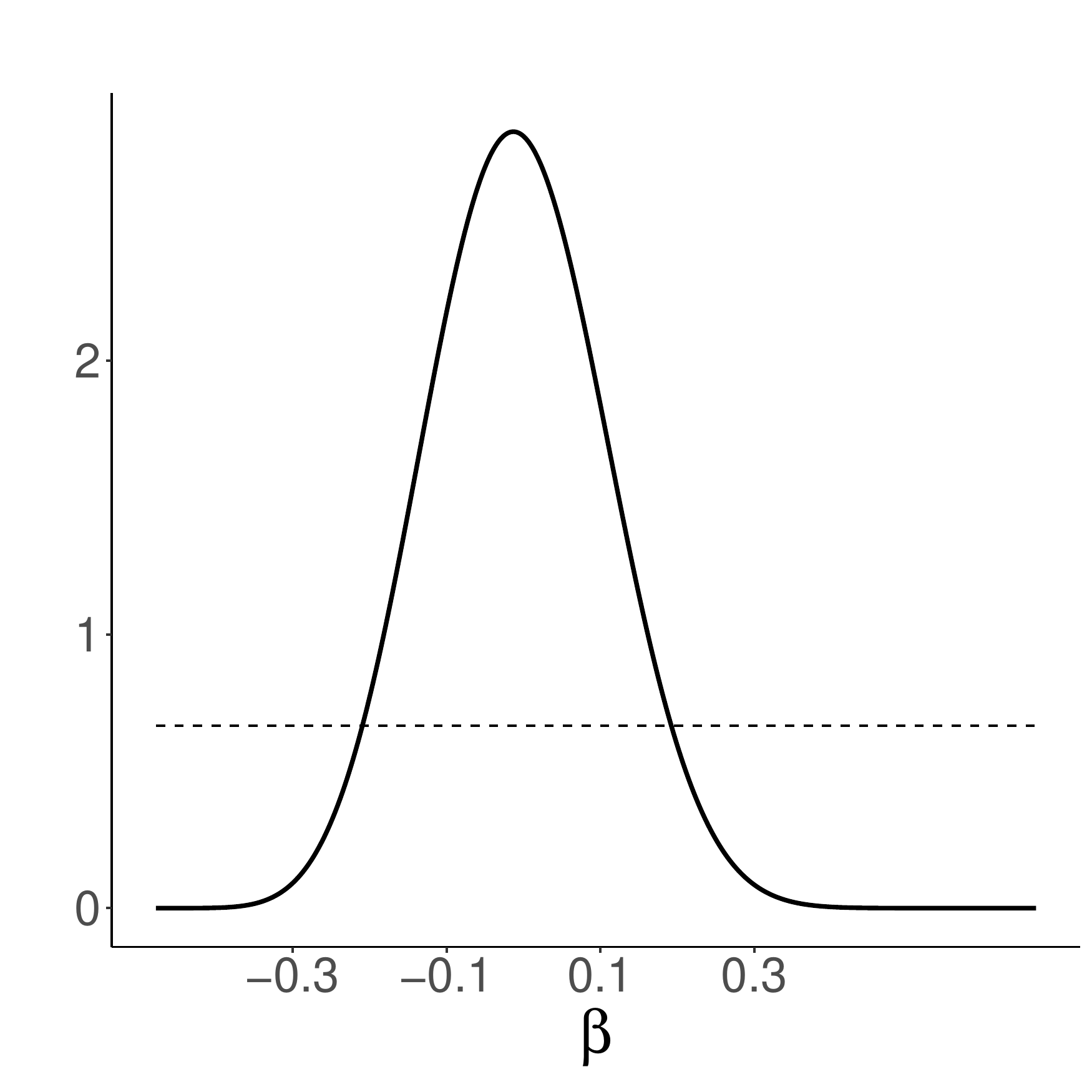}
	\caption{$\widetilde{\pi}_{\LA}(\beta|\mb{Y})$}
	\end{subfigure}
\caption{(a) -- (d) prior (- - -) and approximate posterior (---) distributions for the four parameters for the astronomy data of Example \ref{subsec:astro}.}
\label{fig:astroparam}
\end{figure}

\begin{figure}[t]
\centering
	\includegraphics[height=4in]{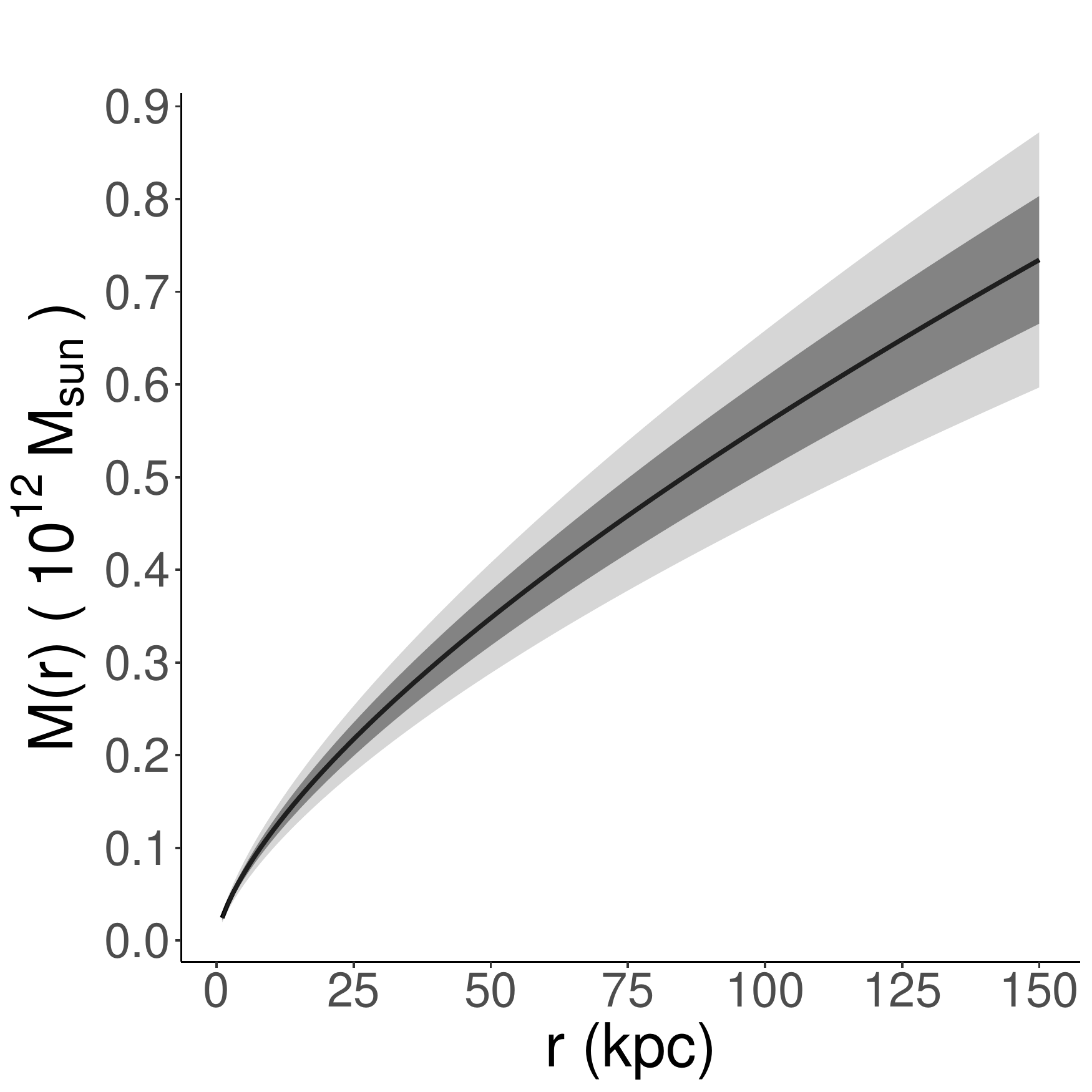}
\caption{Posterior mean mass of the Milky Way Galaxy relative to the mass of the sun as a function of radial distance from Galaxy centre in kiloparsecs (kps) ($M_{\paramsmalllik}(r)$, ---), with one- (dark) and two- (light) standard deviation bands for the astronomy data of Example \ref{subsec:astro}.
}
\label{fig:astromass}
\end{figure}

We fit this model using to Algorithm \ref{alg:implementation}, obtaining results that are broadly comparable to those reported by \citet{gwen2} and \citet{gwen3} accounting for the priors and data inclusion rules used. Figure \ref{fig:astroparam} shows the posterior distributions for $\Psi_{0},\gamma,\alpha,\beta$ and Figure \ref{fig:astromass} shows the estimated cumulative mass profile $M_{\paramsmalllik}(r)$ of the Galaxy for chosen values of $r$. While the posterior mean and standard deviation of $M_{\paramsmalllik}(r)$ are straightforward to obtain using quadrature, an approximation to the posterior density of a nonlinear, trans-dimensional transformation of $\paramsmalllik$ is in general difficult to obtain using our procedure as no algorithm is available to draw samples from $\laplaceapprox(\paramsmalllik|\mb{Y})$, so we instead report the posterior mean along with pointwise one and two standard deviation bands for each value of $r = 1,2,\ldots,150$.

There are several computational challenges in this example which we leave to future work. We do not incorporate missing data into the hierarchical model as done by \citet{gwen3} and \citet{gwen4}. Further, the parameters $\parambig$ and $\paramsmalllik$ are subject to complicated nonlinear constraints implied by the underlying physics, and these constraints depend both on $\paramsmalllik$ and $\Wmode_{\paramsmalllik}$. Advanced methods for optimization in the presence of such nonlinear constraints are readily available (see the \texttt{IPOPT} software of \citet{ipopt} and corresponding \texttt{ipoptr R} package). However, in the presence of measurement errors this would require the derivatives of the constraints which involves differentiating through $\Wmode_{\paramsmalllik}$ with respect to $\paramsmalllik$. This is a challenging task which has recently been investigated in a related context by \citet{hybrid}, and we leave its implementation here to future work.


\section{Discussion}\label{sec:discussion}

We have defined a novel class of Extended Latent Gaussian Models and developed approximate Bayesian inference methodology for this class. The method relies less on sparse matrix algorithms but also depends on matrices of smaller size than previous approaches, and we have provided numerical evidence of our model providing faster run times than \RINLA{} and \MCMC{}, as well as scaling to large sample sizes (\S\ref{sec:comp}). We have proved that the error in our approximation converges to zero as the sample size increases, as long as the assumptions ensuring convergence of the error in each component are satisfied (\S\ref{sec:theory}). Further, we demonstrated three challenging examples of ELGMs and fit them with our procedure: inference for a continuous spatial field using aggregated point process data, including a comparison to \MCMC{} (\S\ref{subsec:aggspatial}); a Cox Proportional hazards model with partial likelihood for mapping the spatial variation in Leukaemia survival times, for which we are unaware of any other method for making Bayesian inferences (\S\ref{subsec:coxph}); and an astrophysical model for estimating the mass of the Milky Way galaxy accounting for multivariate measurement uncertainties (\S\ref{subsec:astro}). The core method (Algorithm \ref{alg:implementation}) is implemented in the open source \texttt{aghq} package in the \texttt{R} language.

There are a number of compelling avenues for future research. While we are unaware of any theoretical results pertaining to the finite-sample performance of Gaussian and Laplace approximations to posterior distributions, several authors \citep{gmrfmodels,inla} have suggested that such approximations may be inaccurate in finite samples, and this merits further attention. \citet{inla} and \citet{simplifiedinla} introduce approximations which are empirically more accurate than the Gaussian, however they work for \emph{marginal} posterior distributions only. In contrast, our procedure depends on access to fast independent sampling from the approximate \emph{joint} posterior (Algorithm \ref{alg:implementation}), which we use to compute complicated posterior summaries in a straightforward manner (\S\ref{subsec:aggspatial}, \S\ref{subsec:coxph}). As noted in the discussion to \citet{inla}, this is a compelling practical advantage of MCMC algorithms, and we believe it is important to retain this advantage when developing methods for approximate Bayesian inference. 

Another potential method for comparison would be Variational Inference (VI) methods based on Gaussian approximations \citep{variationalinference}. Gaussian VI methods find the Gaussian approximation which minimizes a lower bound on the Kullback-Leibler divergence to the true posterior---a complicated, non-convex optimization---in contrast to our approach which matches the mode and curvature of the true posterior. Computationally, our approach differs from VI in that we require only convex optimizations of a tractable objective function, which are therefore computationally feasible (\S\ref{subsec:elgms:computational}), while VI requires a challenging non-convex optimization of an objective function whose value and gradient cannot be calculated analytically, leading to difficulty in implementation. We provide code for fitting the model of \S\ref{sec:comp} with VI using \rstan{}, finding that the model either crashes or provides unreliable answers as determined by the Pareto $\widehat{k}$ diagnostic \citep{yesbutdiditwork,psis,psisloo}, for a variety of sample sizes and priors. While it is possible that an expert user could tune the implementation to produce more reliable answers, our approach does not require such user tuning, again owing to its reliance on stable, tractable convex optimization. We leave further comparisons to VI for future work.

To our knowledge, Theorem \ref{thm:convergence} is the first formal theoretical result pertaining to a nested approximation, of the type made popular in the INLA method of \citet{inla}. In our proof, we explicitly handle the non-standard application of three different types of approximations, including a misspecified Gaussian, a marginal Laplace, and the re-use of adapted quadrature points and weights for approximating two different integrals. It is possible that with further work, this technique could be expanded to apply to other nested approximation algorithms, like the ones proposed by \citet{inla} and \citet{simplifiedinla}.

\section*{Acknowledgements}
The authors are grateful for helpful comments provided by two referees and an associate editor, as well as Blair Bilodeau, Yanbo Tang, and Gwen Eadie. Parts of this manuscript appeared in the first author's PhD thesis for which he is grateful for comments from Gwen Eadie, Linbo Wang, Nancy Reid and Anthony Davison. This work was supported by the Natural Sciences and Engineering Research Council of Canada's Discovery Grant and Postgraduate Scholarships programs.

\section*{Conflict of Interest}
The authors report there are no competing interests to declare.



\bibliographystyle{chicago}

\bibliography{bibliography}
\end{document}